%% file: main.tex
\documentclass[twocolumn]{aastex62}
\newcommand{\ate}{\eta_{\text{\,x/r}}}

\newcommand{\nicer}{\textit{NICER}}
\newcommand{\xmm}{\textit{XMM-Newton}}
\newcommand{\repeater}{FRB\,20220912A}
\newcommand{\meor}{FRB\,20200120E}
\newcommand{\eff}{Effelsberg}
\newcommand{\swift}{\textit{Swift}}
\newcommand{\fluxunits}{erg\,cm$^{-2}$\,s$^{-1}$}
\newcommand{\dmunits}{pc\,cm$^{-3}$}

\newcommand{\anedit}[1]{#1}
\usepackage{bm}
\newcommand{\rthree}{FRB\,20180916B}
\usepackage{amsmath}
\usepackage{physics}

\usepackage{verbatim}
\shorttitle{X-ray Non-Detection of Bursts from FRB 20220912A}
\shortauthors{Cook, A. M. et al.}
\begin{document}
\title{Contemporaneous X-ray Observations of 30 Bright Radio Bursts from the Prolific Fast Radio Burst Source FRB 20220912A}
\correspondingauthor{Amanda M. Cook}
\email{cook@astro.utoronto.ca}
\input{authors.tex}

\begin{abstract}
We present an extensive contemporaneous X-ray and radio campaign performed on the repeating fast radio burst (FRB) source FRB 20220912A for eight weeks immediately following the source's detection by CHIME/FRB. This includes X-ray data from \xmm, \nicer, and \swift, and radio detections of FRB 20220912A from CHIME/Pulsar and \eff. We detect no significant X-ray emission at the time of 30 radio bursts with upper limits on 0.5--10.0 keV X-ray fluence of ${(1.5-14.5)} \times 10^{-10}$ erg cm$^{-2}$ (99.7\% credible interval, unabsorbed) on a timescale of 100 ms. 
Translated into a fluence ratio $\ate = F_{\text{X-ray}}/F_{\text{radio}}$, this corresponds to $\ate < {7 \times 10^6}$. For persistent emission from the location of FRB 20220912A, we derive a 99.7\% 0.5--10.0 keV isotropic flux limit of ${8.8\times 10^{-15}}$ \fluxunits\, (unabsorbed) or an isotropic luminosity limit of ${1.4\times 10^{41}}$ erg s$^{-1}$ at a distance of 362.4 Mpc. 
We derive a hierarchical extension to the standard Bayesian treatment of low-count and background-contaminated X-ray data, which allows the robust combination of multiple observations. This methodology allows us to place the best \anedit{(lowest)} 99.7\% credible interval upper limit on an FRB $\ate$ to date, $\ate  < 2\times10^6$, assuming that all thirty detected radio bursts are associated with X-ray bursts with the same fluence ratio. If we instead adopt an X-ray spectrum similar to the X-ray burst observed contemporaneously with FRB-like emission from Galactic magnetar SGR 1935+2154 detected on 2020 April 28, we derive a 99.7\% credible interval upper limit on $\ate$ of ${8}\times 10^{5}$, which is only \anedit{3} times the observed value of $\ate$ for SGR 1935+2154.
\end{abstract}.


\section{Introduction} \label{sec:intro}
Fast radio bursts (FRBs) are bright, millisecond-duration bursts of unknown astrophysical origin. While we have only detected a single burst from most FRB sources, the first detection of repeat bursts from a single source suggested a non-cataclysmic origin for at least some of these extremely energetic sources \citep{2016Natur.531..202S}. Astronomers have discovered dozens of repeating sources, or \textit{repeaters}, making up roughly 7\% of all published FRBs \citep{2019ApJ...885L..24C, 2020ApJ...891L...6F, catalog, 2023ApJ...947...83C}. 

When a repeater is identified, it can be localized interferometically through follow-up observations and we can coordinate pointed, high-angular-resolution, and sensitive high-energy (HE) observations. Many FRB source theories make predictions for the HE counterparts \citep[or lack thereof, see e.g.,][for a summary]{2019PhR...821....1P}. 
Catching a burst from a repeater is nontrivial; although repeater burst arrival times are clustered temporally, they are still apparently stochastic \citep{2018MNRAS.475.5109O, 2020A&A...635A..61O, 2021MNRAS.500..448C,2023ApJ...944...70G}.

Progenitor theories for repeating FRB sources often invoke connections to neutron stars, in particular pulsars and magnetars, to explain the burst and source properties such as high linear polarization, large Faraday rotation measures, location in star forming regions, coherence, and similar durations, fluences, and waiting times \citep{2013arXiv1307.4924P, 2014ApJ...797...70K, 2015Natur.528..523M, 2017ApJ...843L...8B, 2017ApJ...843L..26B, 2017ApJ...834L...7T,2017MNRAS.468.2726K,2017ApJ...841...14M, 2018Natur.553..182M, 2020Natur.577..190M, 2021AnA...656L..15P}. Neutron-star FRB models typically come in one of two flavors, one in which emission is produced by a synchrotron maser \citep[e.g.,][]{10.1093/mnrasl/slu046, 10.1093/mnrasl/slw202, 2018ApJ...864L..12L,2019MNRAS.485.4091M,Khangulyan_2022}, and one in which the emission is produced near the neutron star magnetosphere \citep[e.g.,][]{2009AstL...35..241E,2014A&A...562A.137F,2016ApJ...823L..28G, 2016ApJ...822L...7W,2017ApJ...838L..13L,Wadiasingh_2019,2023MNRAS.519..497T}. 

In 2020 and again in 2022, FRB-like\footnote{We say ``FRB-like'' since typical FRBs are more than three orders of magnitude more energetic ($ 10^{36} - 10^{41}$ erg) than this Galactic radio burst (3$\times10^{34}$ erg). However, notably, the current closest repeater has produced bursts at this energy scale \citep{2022NatAs...6..393N}.} radio bursts from the Galactic magnetar SGR 1935+2154 were detected at the time of simultaneous X-ray bursts \citep{2020Natur.587...54C, 2020Natur.587...59B, 2021NatAs...5..414K, 2022ATel15681....1D,2022ATel15682....1W,2022ATel15686....1F, 2023arXiv231016932G}, 
adding evidence to the magnetar origin of FRBs by significantly narrowing the energy gap between the two phenomena. The X-ray bursts from SGR 1935+2154, placed at typical FRB distances (hundreds to thousands of megaparsecs), are far too dim to be detected by modern X-ray observatories. However, if one assumes that X-ray burst luminosities scale proportionally to those of their radio counterparts, such an X-ray counterpart to an especially bright FRB should be detectable within a few dozen megaparsecs. 

This and other magnetar observations, and many FRB source theories, provide predictions of X-ray burst luminosities as well as which should be observed first -- the radio or X-ray burst \citep{kaspireview, 2019MNRAS.485.4091M, 2020ApJ...891...82W, 2020ApJ...899L..27M}. As such, X-ray counterparts are sought out as a valuable diagnostic between competing models. 

\subsection{Existing limits on X-ray counterparts of FRBs}
In recent years, astronomers have been able to narrow the luminosity and duration phase space of possible \textit{bona fide}\footnote{`\textit{bona fide}' here, as suggested by \cite{2020ApJ...899L..27M}, means the HE emission associated with the radio bursts rather than the persistent emission or an independent X-ray burst-producing mechanism. This is often referred to as `prompt' in the literature but `\textit{bona fide}' is perhaps more precise since the intrinsic time delay between any X-ray emission and radio emission is unknown.} X-ray counterparts of FRBs. For FRB 20121102A, the first repeater discovered \citep{2016Natur.531..202S}, \cite{r1limits} placed a $5 \sigma$ upper limit of $3 \times 10^{-11}$ erg cm$^{-2}$ on the 0.5--10.0 keV absorbed fluence for X-ray bursts at the time of radio bursts from repeating FRBs for durations $< 700$ ms. This corresponds to an upper limit on the burst energy of $4 \times 10^{45}$ erg  (0.5-10.0 keV) at 972 Mpc \citep[distance from][]{2017ApJ...834L...7T}. This limit is constraining for predictions of the most luminous X-ray FRB-counterpart scenarios, but is still an order of magnitude larger than the X-ray luminosity of the brightest Galactic magnetar giant flare \citep{2005Natur.434.1098H,2005Natur.434.1107P}. 
\cite{2020ApJ...901..165S}, \cite{Pilia_2020} and \cite{2023AnA...676A..17T} place even deeper limits on X-rays at the time of radio bursts from FRB\,20180916B, which is only 150 Mpc away \citep{2017ApJ...834L...7T,2020Natur.577..190M}. By assuming that X-ray bursts of equal fluence are emitted at the time of each radio burst, \cite{2021AnA...656L..15P} derive 2.0--10.0 keV isotropic energy upper limits of $1.1\times10^{44}$ erg for X-ray bursts from FRB\,20201124A, which is consistent with either magnetar scenario.

\cite{2020ApJ...897..146C} summarized all available X-ray luminosity limits on FRB counterparts, and combined the fluence distribution of the FRB population with results from many wide-field untargeted surveys for fast transients, spanning optical to very-HE (TeV) bands. The limits \cite{2020ApJ...897..146C} were able to place on the HE-to-radio fluence using data from wide-field surveys were similar to those previously derived from dedicated/pointed observations.

The nearest repeater discovered to date, FRB\,20200120E, is located in a globular cluster associated with the spiral galaxy M81 \citep{m81r, m81rloc}. The source has a luminosity distance of 3.6 Mpc \citep{m81rloc}, making it the nearest known extragalactic repeater and a premier FRB target for contemporaneous X-ray counterpart detection. \cite{2023arXiv230810930P} did not detect X-ray emission at the time of radio bursts from \meor\,, with isotropic energy upper limits of $\sim 10^{40}$ erg in the 0.5--10 keV range,. This study ruled out X-ray counterparts to radio bursts from \meor\, analogous to magnetar giant flares, as well as some bright magnetar-like intermediate flares and short X-ray bursts. Additionally, \cite{2023arXiv230810930P} ruled out ultraluminous X-ray bursts from \meor, which had been previously detected from unknown sources in extragalactic globular clusters and proposed as a possible source of repeating FRBs \citep{2005ApJ...624L..17S,2013ApJ...779...14J,2016Natur.538..356I,2022ApJ...937....9C}. It is still unknown if FRB\,20200120E is exceptional given its location in a globular cluster, and whether it has a different source type compared to FRB\,20121102A that sits within a star forming region \citep{2017ApJ...834L...7T}.

These deep HE observations have also allowed for sensitive persistent limits to be placed on FRBs. \cite{r1limits, 2020ApJ...901..165S} and \cite{2023arXiv230810930P} place limits on persistent soft ($0.5-10.0$ keV) X-ray luminosity at the location of these FRBs: $3 \times 10^{41}$, $2 \times 10^{40}$, and $9.8\times 10^{36}$ erg s$^{-1}$ for FRBs 20121102A, 20180916B, and 20200120E, respectively. Under the assumption of negligible X-ray absorption local to the source, these limits have ruled out emission similar to the brightest persistent ultraluminous X-ray (ULX) sources\footnote{The persistent X-ray luminosity upper limit of \meor\, is lower than the luminosities of ULXs \citep{2023arXiv230810930P}.} \citep{10.1093/mnras/stab3001, 2023arXiv230703766E}, and a fraction of the brightest high-mass and low-mass X-ray binaries (HMXBs and LMXBs, respectively; \citealt{2023arXiv230810930P}). The \meor\, limits are below the level of persistent emission from a surrounding nebula similar to that of the Crab Nebula, but cannot rule out persistent emission similar to Galactic magnetars, or HMXBs/LMXBs on a population level \citep{2023arXiv230810930P, 2005Natur.434.1098H, 2005Natur.434.1107P, 2008ApJ...685.1114I, 2001ApJ...558L..47K, 2020ApJ...904L..21Y, 2021NatAs...5..408Y, 2023A&A...677A.134N, 2023A&A...675A.199A}.

\subsection{\repeater}
On 2022 October 15, the Canadian Hydrogen Intensity Mapping Experiment Fast Radio Burst Project (CHIME/FRB) announced the discovery of a new, hyperactive, repeating FRB 20220912A \citep{2022ATel15679....1M}. In the days following, nine Astronomer's Telegrams were posted, reporting radio detections spanning almost four octaves, $111-1530$ MHz \citep{2022ATel15693....1R, 2022ATel15699....1H, 2022ATel15713....1F, 2022ATel15716....1R, 2022ATel15720....1R, 2022ATel15727....1K, 2022ATel15733....1Z, 2022ATel15734....1P, 2022ATel15735....1S}. The rate of activity for the source has been measured at nearly 400 bursts per hour at L-band \citep{2022ATel15723....1F, 2022ATel15733....1Z, 2023ApJ...955..142Z} more than a month after its first detection; this was the most active repeater discovered at the time, a record only recently broken by FRB 20240114A (measured as high as $\sim 500$ bursts per hour at L-band; \citealt{2024ATel16420....1S, 2024ATel16505....1Z}). 

FRB\,20220912A is relatively nearby: \cite{2023ApJ...949L...3R} report a spectroscopic redshift of $z=0.077$ based on the candidate host galaxy, PSO J347.2702+48.7066, or a luminosity distance of $362.4$ Mpc assuming a flat cosmology with parameters from \cite{2020A&A...641A...6P}. \cite{2023arXiv231214490H} reported the position of \repeater\, to a precision of a few milliarcseconds, and while they report a continuum radio source coincident with this position on arcsecond scales, they do not find evidence for a persistent radio source associated with \repeater\, (i.e., no compact persistent emission on milliarcsecond scales). 

Using simultaneous KeplerCam and LCO-r band observations, \cite{2022arXiv221103974H} placed a luminosity limit of $\nu L_{\nu} \sim (0.3-1.5) \times 10^{42}$ erg s$^{-1}$ on prompt optical emission from FRB\,20220912A at the time of a radio burst. \cite{2024arXiv240504802P} place a $3\sigma$ upper limit on persistent 0.4 – 30 MeV $\gamma$-ray luminosity of $L_\gamma < 7.1\times10^{43}$ erg s$^{-1}$ for \repeater\, with the AGILE satellite. Also using simultaneous AGILE and Northern Cross radio telescope observations, \cite{2024arXiv240504802P} constrain the prompt radio efficiency at the time of a radio burst, $E_{\gamma}/E_{r} < 1.5\times10^9$ at the 3$\sigma$ level.
 \repeater\, shows evidence of `nanoshots': \cite{10.1093/mnras/stad2847} reported the detection of bursts with fluences exceeding 400 Jy ms that display broadband, narrow ($\sim 16\,\mu$s), bright (peak $\sim 450$ Jy) microstructure. 

The extremely high burst rate suggests that this source is exceptional, and coupled with the fact that it is well localized and nearby, this source offers a unique opportunity to place deep limits on the HE counterpart of a repeating FRB. In this paper, we outline a campaign of simultaneous radio and X-ray observations from CHIME/FRB, \anedit{CHIME/Pulsar}, \swift, the \textit{Neutron star Interior Composition Explorer} (\nicer), \xmm, and the 100-m Effelsberg radio telescope spanning October--December 2022. In Section \ref{sec:data}, we introduce the various instruments and describe our observations. In Section \ref{sec:analysis}, we present constraints on persistent X-ray emission and on the X-ray emission at the time of radio bursts from the source. We also detail our searches and resultant limits on bursts/flares of varying durations at other times during the observation. Also in Section \ref{sec:analysis}, we compute and report an upper limit on prompt X-ray emission,  assuming that there is an X-ray burst at the time of each radio burst, stacking thirty X-ray non-detections at the times of radio bursts to increase sensitivity. In Section \ref{sec:discussion}, we place these limits in the context of previous limits at a range of frequencies, and discuss the implications of these limits on various FRB models. 

\section{X-ray and Radio Data} 

\label{sec:data}
\subsection{CHIME}
CHIME is a transit radio telescope, located at the Dominion Radio Astrophysical Observatory near Penticton, British Columbia, Canada, which operates in the 400--800 MHz frequency range \citep{2022ApJS..261...29C}. It is comprised of four 20\,m~$\times$~100\,m, North--South oriented, cylindrical parabolic reflectors, each of which has 256 dual-polarization feeds. CHIME has an instantaneous field-of-view~(FoV) of more than 200 deg$^{\text{2}}$~\citep{2017ursi.confE...4N}.

CHIME is equipped with multiple backends, each tailored for specific science cases. In this work, we make use of the CHIME/FRB \citep{overview} and CHIME/Pulsar \citep{PSRoverview} backends, which we describe below. 

\paragraph{CHIME/FRB} The realtime pipeline of CHIME/FRB searches 1024 beams for radio pulses with durations of a few to hundreds of milliseconds, such as those produced by pulsars and FRBs. This realtime FRB search is performed on 16k frequency channels at 0.983-ms time resolution. The realtime pipeline consists of four stages. The first stage is L0, which primarily spatially correlates signals, beamforms, and up-channelizes. This is followed by L1, which performs an initial radio frequency interference (RFI) cleaning and searches for dispersed signals via a highly optimized tree dedispersion algorithm called \textit{bonsai}. L2/L3 work together to sift through events to further distinguish between RFI, known sources, new sources, and Galactic versus extragalactic events, and then determines what kind of data should be stored for a given event. Finally, L4 writes and stores metadata headers of the signals \citep{overview}.
\paragraph{CHIME/Pulsar} CHIME/Pulsar is a digital pulsar observing system, which is capable of producing up to 10 digitally-steerable beams formed by the CHIME FX-correlator \citep{PSRoverview}. We used a steerable tracking beam to observe \repeater\ for roughly 21 minutes each day, while the source transited through CHIME's primary beam. Search-mode filterbank data were recorded with a time resolution of 40.96\,$\mu$s and a frequency resolution of 390.625\,kHz. These observations were conducted following the discovery of high activity of \repeater\ in mid-October using CHIME/FRB~\citep{McKinven+2022}. 

Here, we report the radio bursts that were detected with CHIME/Pulsar during good time intervals (GTIs) of our simultaneous X-ray observations with \nicer, \xmm, or \swift. In total, thirty radio bursts occurred during our simultaneous X-ray exposures, twenty six ocuured during the \nicer\, observations. The properties of the radio bursts are provided in Table~\ref{tab:psr}; all properties were derived assuming a fiducial dispersion measure (DM) of 219.456 \dmunits, which was calculated by maximizing the signal-to-noise ratio (S/N) of a bright, broadband burst \citep{McKinven+2022}. \anedit{The times of arrival which we report are referenced to an infinite frequency and translated to the barycenter of the solar system. We performed the barycentric correction using the} \verb|pintbary| \anedit{tool in the} \verb|pint| \anedit{software package} (version 0.9.7, observatory option \verb|chime| and using JPL planetary ephemeris DE405; \citealt{2021ApJ...911...45L, 1998A&A...336..381S}).

Previous observations indicate that the assumed system temperature of \anedit{CHIME/Pulsar} is overestimated by a factor of 2–3, leading to calculated fluxes being underestimated by the same factor \citep{Good_2021}. Hence we report only lower limits on fluence the radio bursts detected by CHIME/Pulsar. Two of the bursts \anedit{(B21 and B27)} in our sample were co-detected by CHIME/Pulsar and CHIME/FRB, and had voltage data saved from the latter \citep[][]{2021ApJ...910..147M}. In these two cases, we derive more reliable fluence measurements and we hence report the estimates, along with 1$\sigma$ uncertainties on these measurements. 

\begin{deluxetable*}{cccc}
\caption{\label{tab:psr} Radio burst detections during X-ray observations.}
\tablehead{
\colhead{Burst Number} & \colhead{Barycentric ToA\tablenotemark{a}} &\colhead{Fluence}& \colhead{Observation ID}\\ &
\colhead{(MJD)} & \colhead{(Jy ms)}& \colhead{}}
\startdata
CHIME/Pulsar && & \swift \\ \hline
B1 & 59867.2337651529(5) & $>1.12$ & 00015380001\\
B2 & 59867.2361652994(5) & $>0.52$ & 00015380001\\
B3 & 59868.2302000210(5) & $>0.24$ & 00015380002\\\hline
CHIME/Pulsar && & \nicer \\ 
\hline
B4& 59880.1990692887(5) &  $> 1.27$ & 5203470102 \\
B5& 59880.2007281066(5) &  $> 0.32$ & 5203470102 \\
B6& 59880.2012657886(5) &  $> 0.45$ & 5203470102 \\
B7& 59880.2014033844(5) &  $> 0.42$ & 5203470102 \\
B8& 59880.2021430438(5) &  $> 0.42$ & 5203470102 \\
B9& 59880.2021439275(5) &  $> 0.56$ & 5203470102 \\
B10& 59880.2039849901(5) &  $> 0.91$ & 5203470102 \\
B11& 59880.2045395375(5) &  $> 1.13$ & 5203470102 \\
B12& 59882.1909163952(5) &  $> 1.97$ & 5203470103 \\
B13& 59882.1915613202(5) &  $> 0.67$ & 5203470103 \\
B14& 59882.1951750479(5) &  $> 0.66$ & 5203470103 \\
B15& 59882.1951762690(5) &  $> 0.70$ & 5203470103 \\
B16& 59882.1951801563(5) &  $> 0.83$ & 5203470103 \\
B17& 59882.1951805052(5) &  $> 0.43$ & 5203470103 \\
B18& 59882.1955737831(5) &  $> 0.79$ & 5203470103 \\
B19& 59884.1880391532(5) &  $> 0.78$ & 5203470104 \\
B20& 59884.1902762081(5) &  $> 0.36$ & 5203470104 \\
B21& 59884.1908769493(5) &  $ 12.3\pm1.4$ & 5203470104 \\
B22& 59886.1846954935(5) &  $> 0.12$ & 5203470105 \\
B23& 59886.1892703034(5) &  $> 0.85$ & 5203470105 \\
B24& 59889.1707981026(5) &  $> 2.02$ & 5203470107 \\
B25& 59889.1720946333(5) &  $> 0.66$ & 5203470107 \\
B26& 59889.1738768160(5) &  $> 0.65$ & 5203470107 \\
B27& 59889.1742898498(5) &  $13.6\pm 1.5$ & 5203470107 \\
B28& 59889.1748964928(5) &  $> 0.54$ & 5203470107 \\
B29& 59889.1749662400(5) &  $> 1.69$ & 5203470107 \\ \hline
Effelsberg&&&  \xmm \\ 
\hline
B30 & 59922.706875601616 &  $12.6 \pm 0.8$ & 0903220101\\
\enddata
\tablenotetext{a}{Burst time of arrival~(ToA) at infinite frequency, after correcting for dispersion and translating to the barycenter of the solar system.}
\end{deluxetable*}

\subsection{Effelsberg}
    The Effelsberg Radio Telescope is a 100-m parabolic dish radio telescope located near Bonn, Germany, operated by the Max Planck Institute for Radio Astronomy \citep{2011JAHH...14....3W}.

We used the P210-7 receiver, which is a seven-beam, cryogenically-cooled receiver system with a 
400 MHz bandwidth, centered at 1400 MHz. We performed three observations of the source, spanning 18 hours total. The data quality was confirmed based on observations of a bright pulsar, PSR B0355+54. Unfortunately, the observations were taken during a period of high RFI at the \eff\, site. 
We detected only one burst from the source during these observations, which occurred simultaneously during our \xmm\, exposures. The burst had an observed duration of 10.88\,ms and a S/N of 152.9 when de-dispersed to 220 \dmunits. \anedit{This is slightly different from the DM assumed in the CHIME/Pulsar analysis, as the Effelsberg search was completed before the detection of the bright burst we based our 219.456 \dmunits\ estimate on.} We calculate a burst fluence of 12.6 Jy ms assuming a system equivalent flux density of 15 Jy, using an emitting bandwidth of 180 MHz, and accounting for the dual-polarizations used in calculating the S/N. \anedit{The time of arrival of the burst detected by Effelsberg reported in Tables \ref{tab:psr} and \ref{tab:paulaaron} is referenced to an infinite frequency and converted to the barycenter of the solar system. We performed the barycentric correction to the topocentric arrival time using the Astropy package \citep{2022ApJ...935..167A}. For Effelsberg’s geographic coordinates, we used longitude 6.882778 degrees and latitude 50.5247 degrees.}

\subsection{\textit{XMM-Newton}} 

Launched in 1999 by the European Space Agency, \xmm\ is a powerful X-ray space observatory. \xmm\, has three identical mirror modules consisting of 58 nested, grazing incidence mirrors. Each mirror module has a focal length of up to 7.5\,m; for imaging, \xmm\, has an effective collecting area of $900$ cm$^{2}$ at 7 keV and a FoV of approximately 30 arcminutes. We triggered an Anticipated Target of Opportunity observation using one of the three scientific instruments onboard \xmm, the European Photon Imaging Camera (EPIC) \citep{2001A&A...365L..18S}. EPIC is composed of three cameras, the pn-CCD and two MOS-CCD detectors, which can detect X-ray photons in the $0.1-15$ keV energy range. We use EPIC-pn in Large Window mode with the medium filter as our primary instrument. MOS1 and MOS2 data were taken in Partial Window 3 mode with the medium filter. 

The data were reduced using SAS version 20.0.0, using tools \verb|epproc| and \verb|emproc| with default settings, which select for exposure, CCD, attitude, GTIs, bad pixels, and filters. We applied a barycentric correction using the known source position and the SAS tool \verb|barycen|. Additionally, for the persistent limit, we form the GTIs where the average 0.4--10 keV photon count rate across the image was less than 0.4 photons per second. This threshold is suggested by the \xmm\, team in their analysis threads\footnote{\url{https://www.cosmos.esa.int/web/xmm-newton/sas-threads}} to exclude intervals of flaring particle background. 

\begin{figure*}
    \centering
    \plotone{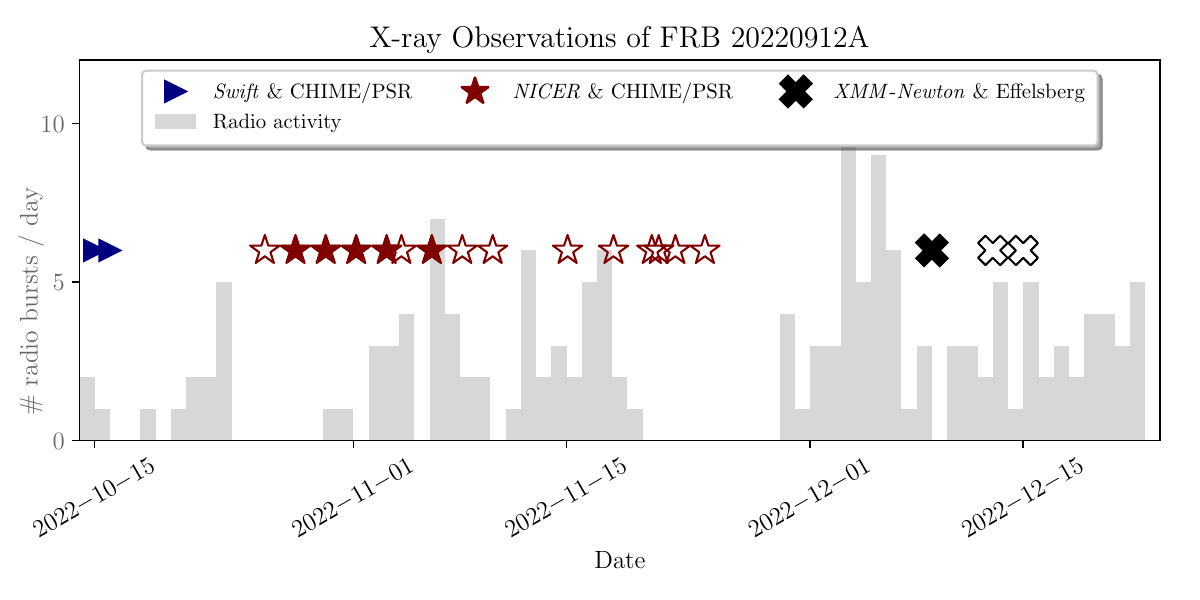}
    \caption{Timeline of X-ray observations compared with the number of bursts detected by CHIME/FRB per few-minute daily source transit (gray histogram, not corrected for exposure).  The observations are summarized in more detail in Section \ref{sec:data}. The blue triangles and red stars represent \swift\, and \nicer\, observations, respectively, for which CHIME/Pulsar provided simultaneous radio coverage.  The black crosses denote our \xmm\, observations with contemporaneous \eff\, radio coverage.  Filled markers denote observations with radio bursts detected during a GTI during an X-ray observation. These X-ray data were obtained during periods of high radio activity from the source.}
    \label{fig:x-raytimeline}
\end{figure*}

Our X-ray observations, along with the number of detected bursts per transit from CHIME/FRB, are shown in Figure \ref{fig:x-raytimeline} as an indicator of source activity at the time of our X-ray data. These burst rates have not been corrected for exposure nor are they necessarily complete (a robust database search, which would include a clustering algorithm, has not been conducted). For other studies that make statements about the source's radio properties rather than only the X-ray emission, this should be taken into account. A full catalog of CHIME radio bursts detected from the source will be presented elsewhere. 

\subsection{\textit{NICER}}
\textit{NICER} is an X-ray telescope, originally designed to study the properties of neutron stars through soft X-ray timing~\citep{2016SPIE.9905E..1HG}. Launched on 2017 June 3, \nicer\, is mounted on the International Space Station. \nicer\ is equipped with an X-ray Timing Instrument (XTI) that consists of 56 X-ray detectors (52 operational on orbit), which provide a peak effective area of $\sim$1500\,cm$^{\text{2}}$ at $\sim$1.5\,keV~\citep{2014SPIE.9144E..20A}. The XTI covers an energy range 0.2--12 keV~\citep{2012SPIE.8443E..13G}. Photons detected by \nicer\ are time-tagged relative to GPS and are accurate to better than 100\,ns root-mean-square (\citealt{LaMarr+2016, Prigozhin+2016}).

We carried out X-ray observations of \repeater\ with \nicer\ between 2022~October~26 and 2022~November~11. High time resolution radio observations were simultaneously performed using one of CHIME/Pulsar's tracking beams during our observational campaign \citep{PSRoverview}.

The \nicer\, observations, as well other X-ray observations performed using other X-ray telescopes, are summarized in Table \ref{tab:xrayobs}.

The \nicer\, data were reduced with the instrument specific software suite \verb|NICERDAS| version 10, included in the HEASoft (version 6.31) software package \citep{2014ascl.soft08004N}. The data were first calibrated and screened using standard NICER-recommended processes in the \verb|nicerl2| task. GTIs were computed with \verb|nimaketime| and events that fall in those GTIs were extracted using \verb|niextract-events|.\anedit{We then used the task} \verb|nicerl3-spect| \anedit{to generate the required ancillary response files for each observation in the NICER-recommended way. The ancillary response files allow us to convert the number of counts from the detector to physical flux units.} Finally, we applied barycenter corrections to the event lists and GTIs using \verb|barycorr|, the known position of the source \citep{2023arXiv231214490H} and JPL planetary ephemeris DE405 \citep{1998A&A...336..381S}. All assumed source properties are summarized in Table \ref{tab:params}.

\begin{deluxetable*}{clcccc}
\caption{\label{tab:xrayobs} X-ray Observations.}
\tablehead{
\colhead{Telescope}& \colhead{Observation ID}& \multicolumn{2}{c}{Start time}& \colhead{End time}&  \colhead{Exposure}\\
\colhead{}&\colhead{}&\colhead{(MJD)}&\colhead{(UTC)}&\colhead{(UTC)}&  \colhead{(s)}}
\startdata
\swift & 00015380001 & 59867.22433
& 2022-10-15T05:23:02  & 2022-10-15T05:42:30  & 895 \\
\swift & 00015380002 & 59868.22155
&  2022-10-16T05:19:02 	& 2022-10-16T05:38:08  & 979 \\
\nicer & 5203470101 & 59878.18130 & 2022-10-26T04:21:04 & 2022-10-26T04:51:00 & 923 \\
\nicer  & 5203470102 & 59880.19297 & 2022-10-28T04:37:53 & 2022-10-28T04:56:20 & 934 \\
\nicer  & 5203470103 & 59882.18068& 2022-10-30T04:20:11 & 2022-10-30T04:58:20 & 2186 \\
\nicer  & 5203470104 &	59884.18132& 2022-11-01T04:21:06 & 2022-11-01T04:56:40 & 2029 \\
\nicer  & 5203470105 &59886.18127 & 2022-11-03T04:21:02 & 2022-11-03T04:59:00 & 2174 \\
\nicer & 5203470106 & 59887.15162 & 2022-11-04T03:38:20 & 2022-11-04T04:13:00 & 1951 \\
\nicer & 5203470107 & 59889.14861 & 2022-11-06T03:34:00 & 2022-11-06T04:13:20 & 2296 \\
\nicer & 5203470108 & 59891.14789 & 2022-11-08T03:32:58 & 2022-11-08T04:14:20 & 2406 \\
\nicer  & 5203470109 & 59893.14745 & 2022-11-10T03:32:20 & 2022-11-10T04:14:40 & 2466 \\
\nicer  & 5203470110 &59898.06802 & 2022-11-15T01:35:42 & 2022-11-15T04:59:19 & 1833 \\
\nicer  & 5203470111 &59901.08515 & 2022-11-18T01:56:36 & 2022-11-18T04:05:13 & 1798 \\
\nicer  & 5203470112 &59903.60396 & 2022-11-20T14:24:42 & 2022-11-20T14:35:54 & 372 \\
\nicer  & 5203470113 & 59904.05746 & 2022-11-21T01:16:40 & 2022-11-21T16:54:51 & 1306 \\
\nicer  & 5203470114 & 59905.1536 & 2022-11-22T03:35:40 & 2022-11-22T03:45:01 & 225 \\
\xmm &0903220101 &59922.41864 &  2022-12-09T10:02:51	& 2022-12-09T19:12:51 & 26863	 \\
\xmm & 0903220401& 59926.40576 & 2022-12-13T09:44:18	& 2022-12-13T18:54:18 &  26857	\\
\xmm &0903220501 & 	59928.38846&  2022-12-15T09:19:23 &	2022-12-15T18:46:03 & 29526	
\enddata

\end{deluxetable*}
\subsection{Swift}
On 2022 October 14, the CHIME/FRB team sent a Target of Opportunity (ToO) request to \swift\ for two observations of \repeater\ during the source's 15 minute transit over CHIME. \swift\ performed observations on 2022 October 15 and 16, for a total exposure of 1.9 ks. \swift's X-ray Telescope (XRT) is a CCD imaging spectrometer, sensitive to 0.2--10 keV photons. In photon counting mode, which was requested for our ToO as the count rate was expected to be very low, XRT has a time resolution of 2.5\,s. During the first observation on 2022 October 15 (Observation ID 00015380001), the Earth limb began encroaching into the field of view about halfway through the observation. This increased the background rate substantially, which caused the ring buffer to overflow/saturate and hence we do not have an accurate upper limit on flux during the second half of the observation. Limits are set using the cleaned XRT event files provided by \swift\, and the \swift\, software tool, \verb|xrtmkarf|, included in the HEASoft software package. \anedit{Photon arrival times are barycentered using the} \verb|barycorr| \anedit{tool, also included in the HEASoft software package, with JPL planetary ephemeris DE405 \citep{1998A&A...336..381S}. }

\section{Analysis}
\label{sec:analysis}
\subsection{Persistent X-ray Emission}
We detect no persistent source consistent with \repeater's position \citep{2023arXiv231214490H} in the \xmm\ EPIC data (within a 90\% containment region, which is 680 pixels or 35 arcseconds). In order to determine the significance of a collection of X-ray photons or lack thereof, we use the methodology presented by \cite{kbn}, who derived a Bayesian expression for confidence intervals in error analysis for photon counting experiments with low numbers of counts. This formalism is often chosen by X-ray astronomers because of its straightforward application to circumstances with non-zero expected background counts. We make use of the Python implementation in the \verb|pwkit| library \citep{2017ascl.soft04001W} in order to compute these statistics. 
We estimate a 99.7\% credible interval (chosen as it produces physically relevant constraints while still being conservative, 99.7\% is $\sim 3\sigma$ Gaussian equivalent) on the 0.5--10.0 keV source count rate from the EPIC pn of between 0 and $5 \times 10^{-4}$ counts per second for the $\sim$ 44 ks GTIs within our observations of the source. \anedit{This 99.7\% credible interval upper limit count rate assumes an average background rate of $3\times10^{-3}$ counts per second, which was estimated from the same observations in a spatial region away from FRB 20220912A and without any obvious X-ray sources.} Hence we infer a 99.7\% credible interval upper limit on unabsorbed flux of \anedit{
${8.8}\times 10^{-15}$ ergs cm$^{-2}$ s$^{-1}$ 
in the 0.5--10 keV range, assuming isotropic emission with ${\Gamma=2}$ power-law spectrum, absorbed by a $N_H = 10^{22}$ cm$^{-2}$ neutral hydrogen column}. Using the luminosity distance of 362.4 Mpc, this corresponds to a 0.5--10.0 keV isotropic-equivalent luminosity of \anedit{${L_X < 1.4 \times 10^{41}}$ ergs s$^{-1}$}.
\anedit{The correct $N_H$ to assume is not obvious. X-rays could be significantly absorbed by intervening material along the line-of-sight of our source. From our own Galaxy, \cite{2016AA...594A.116H} estimate a neutral hydrogen column of $1.42 \times 10^{21}$ cm$^{-2}$ along the line of sight of FRB 20220912A. This can be an estimate of the minimal possible total $N_H$ along the line of sight of the source.  Following the argument in \cite{2020ApJ...901..165S}, the low extragalactic DM\footnote{\anedit{the DM above what is maximally predicted free electron models of the disk of the Milky Way  (MW) along that line of sight, not accounting for the halo, see e.g. \cite{2023ApJ...946...58C}.}} of 94.3 or 97.3 pc cm$^{-3}$ (using the free electron models in NE2001, YWM16 respectively, \citealt{ne2001, ymw16}) suggests there are not orders of magnitude more ionized plasma along the line of sight than what is contributed by our Galaxy. \cite{2013ApJ...768...64H} predict ~$(3\pm 1)\times10^{21}$ cm$^{-2}$ from this extragalactic DM, after accounting for the uncertainty in the relationship and the differing extragalactic DM predictions. To be conservative, we assume an $N_H$ of $10^{22}$ cm$^{-2}$ throughout. Of course, this makes the assumption that the ratio of atomic metals to free electrons is similar to that found in the MW interstellar medium along the entire line of sight. This may not be the case local to the FRB source if, for example, the source was located in a decades-old supernova remnant \citep{2017ApJ...841...14M}. If there was extreme X-ray absorption local to the source, one could expect a $N_H$ value as high as $10^{24}$ cm$^{-2}$ \citep{r1limits}, although such a scenario has been disfavoured for at least some FRB sources, given their detection at low ($\sim 300$ MHz) frequencies \citep{2022ApJ...927...35C,2020ApJ...901..165S}. FRB 20220912A has been detected at frequencies as low as 111 MHz, so the same argument can be applied to this source \citep{2023ARep...67..970F}. The impact of an extreme X-ray absorption on resultant X-ray limits is explored more in \cite{r1limits, 2020ApJ...901..165S}, and from their calculations, we can assume in this scenario our limits would be about an order of magnitude less constraining. }

\swift-XRT also did not detect significant persistent source within the 90\% containment region of \repeater \, (a circle with a 20 pixel or 47 arcsecond radius), which places an upper bound of the 99.7\% credible interval on absorbed source flux between 0.2 and 10.0\,keV of $8.5 \times 10^{-14}$ erg\,cm$^{-2}$\,s$^{-1}$. \anedit{Again, this assumes an average background rate, here $1\times10^{-3}$ counts per second, which we estimate from a spatial region away from FRB 20220912A without any obvious X-ray sources in it.} We report this less constraining value along with the deep limit from \xmm\ EPIC as the \swift-XRT limit was obtained much closer to the source's initial activation, and hence could still be constraining for decaying emission models. 

We do not estimate the \nicer\ persistent upper limit as we detected another faint source with \swift-XRT\ (0.2--10.0 keV X-ray flux $\sim 6\times 10^{-14}$\,\fluxunits\ using the On-Demand XRT products webtool from \citealt{2009MNRAS.397.1177E}) within the FoV of the \nicer \ observations (albiet off-axis) and \nicer\, is non-imaging. The faint source is not consistent with \repeater\, as it is outside of \swift-XRTs 90\% PSF containment region around \repeater. 

We searched a two degree radius around \repeater\, for any gamma-ray source in the \textit{Fermi}-LAT 14-Year Point Source Catalog \citep[4FGL-DR4;][]{2023arXiv230712546B,2022ApJS..260...53A}; but there are no known cataloged gamma-ray sources within this sky region.

\subsection{Prompt X-ray Emission at the Times of Radio Bursts}

\paragraph{\xmm} We detected one radio burst during our \xmm\,observations. The burst occurred during a time of high X-ray background, presumably from soft proton flares. We place a 99.7\% credible region upper limit of 5.8 counts \anedit{given the absence of photons detected by EPIC-pn (using \citealt{kbn})}, or \anedit{$< {1.5 \times 10^{-10}}$ erg cm$^{-2}$} on 0.5--10.0 keV unabsorbed X-ray fluence at the time of the radio burst (having corrected for the DM delay). \anedit{The background rate for each of the short duration/burst upper limits is estimated by averaging the light curve over the 200s surrounding the radio burst ToA, with a buffer equal to the assumed duration of the X-ray burst. For \xmm\ and \swift , we average over the light curve of the 90\% spatial containment region of \repeater, whereas for \nicer , which is non-imaging, we average over the light curve of the entire field.} The upper limit value also assumes a 100 ms X-ray burst duration, photo-electrically absorbed by \anedit{a moderate neutral hydrogen column in that direction (${N_H = 10^{22}}$ cm$^{-2}$)}, assuming a 10\,keV blackbody spectrum. A timescale of 100 ms was chosen as it is comparable to the duration of the X-ray counterpart to the SGR 1935+2154 April 2020 FRB-like radio burst \citep{Mereghetti_2020}. There are no photons within $\pm$ 820 ms of the burst (after correcting for the dispersive delay), hence a similar limit can be placed for bursts of durations up to 820\,ms. 

\paragraph{\nicer} We detected 26 radio bursts with CHIME/Pulsar while \nicer\, was simultaneously observing the source. Of these bursts, and considering the trials factor of 26, the smallest individual 99.7\% credible region upper limit constrains a prompt 0.5--10.0 keV X-ray burst fluence to 9.0 counts, or $< {4.0} \times 10^{-10}$ erg cm$^{-2}$ assuming the same burst properties as for \xmm\, above. The depth of the limit, for a fixed telescope and source distance, is predominantly based on the average background rate at the time the limit is placed. In placing the best limit, we are essentially selecting for the lowest background, and hence we must incorporate a trials factor to reflect the increased probability of observing a low background count rate realization compared to the true background count rate among many trials. The limit reported above is corrected for 26 trials via the Dunn–Šidák correction\citep{doi:10.1080/01621459.1967.10482935}, a simple method to control for the family-wise error rate in multiple hypothesis testing which is conservative for tests that are positively dependent. The Dunn–Šidák correction widens the confidence intervals for a given significance threshold ($\alpha \in [0,1]$) to $100\dot(1-\alpha)^{1/m}$ where $m \in \mathbb{N}$ is the number of hypotheses/trials being tested.  

\paragraph{\swift} We detected three radio bursts with CHIME/Pulsar while \swift\, was simultaneously observing the  source. Of these bursts, and considering the trials factor here of three, the smallest individual 99.7\% credible region upper limit constrains a prompt 0.5--10.0 keV X-ray burst fluence to 6.9 counts, or $< {1.4} \times 10^{-9}$ erg cm$^{-2}$ assuming the same burst properties as for \xmm\, and \nicer\, above.

\subsection{Bursts/Flares of Varying Durations at Other Times During the Observations}

We carried out an untargeted search for bursts of varying durations at times other than at the time of radio bursts during the observations. We searched for significant signals above background from X-ray photons coming from the location of \repeater\, (i.e., within the 90\% PSF-containment regions of \swift-XRT and \xmm\, or any \nicer\ photons, as it is non-imaging). The search was performed by binning the data according to burst width and then checking if the number of photons in a given bin was statistically significant at the 99.7\%  level according to the statistics presented by \cite{kbn}. We used an estimate of the background by averaging the count rate nearby in time to the bin of interest (with a few-bin padding on either side in case a hypothetical burst arrived between two neighboring time bins). We then correct for the problem of multiple hypotheses (or the look-elsewhere effect) with the Dunn–Šidák correction \citep{doi:10.1080/01621459.1967.10482935}. The assumed trials factor is equal to the sum of the number of bins checked for significant signal over all tested bin widths.
We searched for bursts with durations of 1\,ms, 10\,ms, 100\,ms, 1\,s and 10\,s. We find no significant bursts across all X-ray instruments and observations (99.7\% credible region), and we would have been sensitive to bursts at these timescales with $0.5-10.0$ keV fluences $\sim 10^{-9}$ erg cm$^{-2}$. 

\subsection{Stacked Prompt Search}
\label{sec:stacked_section}
\paragraph{Previous methodologies}
While the existing formalism of \cite{kbn} allows us to place limits on X-ray emission at the time of each radio burst, the authors do not provide a formalism for the case where multiple independent measurements are taken of a given source. \cite{r1limits} and \cite{2021AnA...656L..15P} combine the information from multiple independent trials, that is, multiple non-detections of X-ray emission at the time of radio bursts, by assuming that an X-ray burst of the same energy is emitted at the time of each detected radio burst. We derive the Bayesian expression for this calculation in Appendix \ref{app:rate}. This model assumes that the independent observations of $N_i$ X-ray counts at the time of a radio burst is described by a Poisson model with rate parameter $\lambda_i = S + B_i$. For $n$ total X-ray observations, $i \in 1, \ldots n$ is an index describing the $i^{\text{th}}$ X-ray observation.  $B_i$ are the average background rates in each observation and $S$ is the average source count rate during the bursts. The $B_i$ are treated as known and constant. That is, 

\begin{align*}
    N_i \sim \text{Poisson}(\lambda_i = S + B_i). 
\end{align*}
\paragraph{New methodology}
All bursts having the same luminosity is a strong assumption, however, given that most astrophysical transient phenomena that we see can be characterized by some luminosity distribution. 

A more conservative assumption is that an X-ray burst of the same relative X-ray-to-radio fluence is emitted at the time of each radio burst. That is, we ask the question: if one expects to see an X-ray burst whose luminosity scales proportionally to that of a simultaneous radio burst, what is the limit that can be placed? We derive a hierarchical Bayesian expression for the X-ray-to-radio fluence ratio\footnote{We use the symbol $\ate$ rather than simply $\eta$ to differentiate between X-ray-to-radio fluence ratio and radio-to-X-ray fluence ratio respectively. The latter is slightly more common, however the definition of relative fluence ratio is not widely standard in FRB applications. $\ate$ has the benefit of being defined when there is no X-ray counterpart, hence our selection.} $\ate= F_{\text{X-ray}}/F_{\text{radio}}$ in Appendix \ref{app:eta}. $F_{\text{X-ray}}$ is the X-ray fluence (in erg cm$^{-2}$), $F_{\text{radio}}$ is the radio fluence (converted from Jy ms to erg cm$^{-2}$ Hz$^{-1}$ and then multiplied by the emitting bandwidth of the bursts, which can be a conservative underestimate due to our finite observing band). This hierarchical model assumes the following distributions

\begin{align*}
\text{Step I} &  &  N_i & \sim \text{Poisson} (\lambda_i = S_i + B_i) \\
  \text{Step II} & & S_i & \sim \mathcal{N}_0^\infty \left( \frac{\ate F_{\text{radio}, i}}{(\text{Flux/S})} , \frac{\ate \sigma_{F_{\text{radio}, i}}}{(\text{Flux/S})} \right) 
\end{align*}
where $F_{\text{radio, }i}, \sigma_{F_{\text{radio, }i}}$ are the radio fluences of the detected simultaneous bursts and their associated uncertainties, respectively. (Flux/S)$\in \mathbb{R}^+$ is a conversion parameter to turn the X-ray count rates into fluences. This value depends on the underlying spectral model assumed and the effective area of the X-ray telescope, but can be computed using standard X-ray tools. $\mathcal{N}_0^\infty (\mu, \sigma)$ denotes a normal distribution truncated on the left at $0$ with mean $\mu \in \mathbb{R}^+$ and standard deviation $\sigma \in \mathbb{R}^+$. Poisson$(\lambda)$ denotes the Poisson distribution with rate parameter $\lambda \in \mathbb{R}^+$. With CHIME/Pulsar alone, we have only a lower limit on burst fluence, but co-detections between CHIME/Pulsar and other telescopes suggest a factor of 2--3 underestimate. Hence, for these bursts, we instead model the flux distribution in Step II as  $\mathcal{N}_{S_{\text{radio, }i}}^\infty (2S_{\text{radio, }i},S_{\text{radio, }i})$ where $S_{\text{radio, }i} = \ate F_{\text{radio, }i}/(\text{Flux/S})$. This enforces that our reported fluence limits are strict lower limits, and conservatively accounts for the underestimate factor of 2--3. The resulting limits are conservative because they, on average, underestimate the fluence of the CHIME/Pulsar radio bursts, and hence inflate the upper limit we place on $\ate$. A numerical implementation of these models in python and a minimal working example \anedit{is available on Zenodo at} \url{https://doi.org/10.5281/zenodo.12785591} \citep{cook202412785591}.

\begin{figure*}   
\plotone{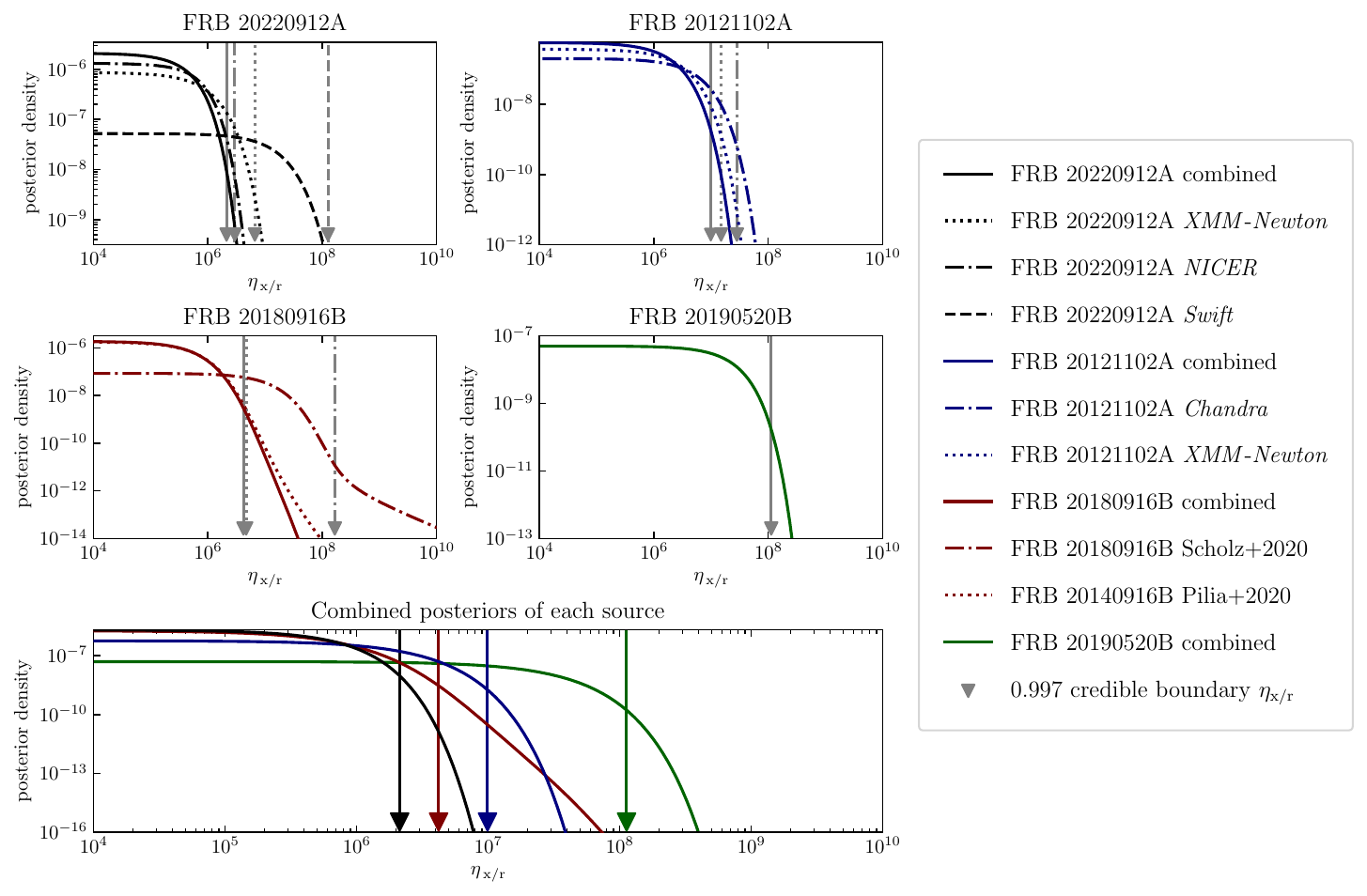}
    \caption{Posterior distributions of $\ate$ (band-integrated X-ray-to-radio fluence ratios) from a selection of soft X-ray observations of FRBs. In the top four panels, the value of $\ate$ corresponding to the upper boundary on the 0.997 credible interval is shown with a gray arrow using the same linestyle. The posterior distributions approach a constant value from the lower $\ate$ bound of the plot to $\ate=0$ and approach zero density as $\ate$ approaches infinity. \textit{Top-left panel:} Posterior distributions on $\ate$ using the Bayesian stacking method described in Section \ref{sec:stacked_section} and Appendix \ref{app:eta}, based on all observations reported in this paper (solid black line). The posterior distribution of $\ate$, computed from simultaneous \xmm\ and Effelsberg data at the time of B30, is shown as a black dotted line. The dot-dashed line shows the posterior distribution of $\ate$, computed from simultaneous \nicer\ and CHIME/Pulsar bursts (B4--B29). The dashed line shows the posterior distribution of $\ate$, computed from the simultaneous \swift\ and CHIME/Pulsar bursts (B1--B3).  \textit{Top-right panel:} Posterior distributions of four radio bursts from FRB 20121102A simultaneously observed with \textit{Chandra}, reported by \cite{r1limits}, and combined using our method (solid blue line). \textit{Middle-left panel:} Same as in the top panels, but instead using the observations reported by \cite{2020ApJ...901..165S} (dotted maroon line) and \cite{Pilia_2020} (dot-dashed maroon line), along with our combined Bayesian stacking measurement (solid maroon line). \textit{Middle-right panel:} Same as in the top panels, but instead based on the observations of 
    FRB 20190520B by \cite{2024arXiv240212084Y} and combined using our Bayesian stacking method (solid green line). 
    \textit{Bottom panel:} Cumulative posterior distributions of $\ate$, combined using our Bayesian stacking method and plotted source-by-source, for comparison. Arrows, colored by source, denote upper limits corresponding to 0.997 $\ate$ credible regions.}
    \label{fig:posteriors}
\end{figure*}

\begin{figure*}
    \centering
\plotone{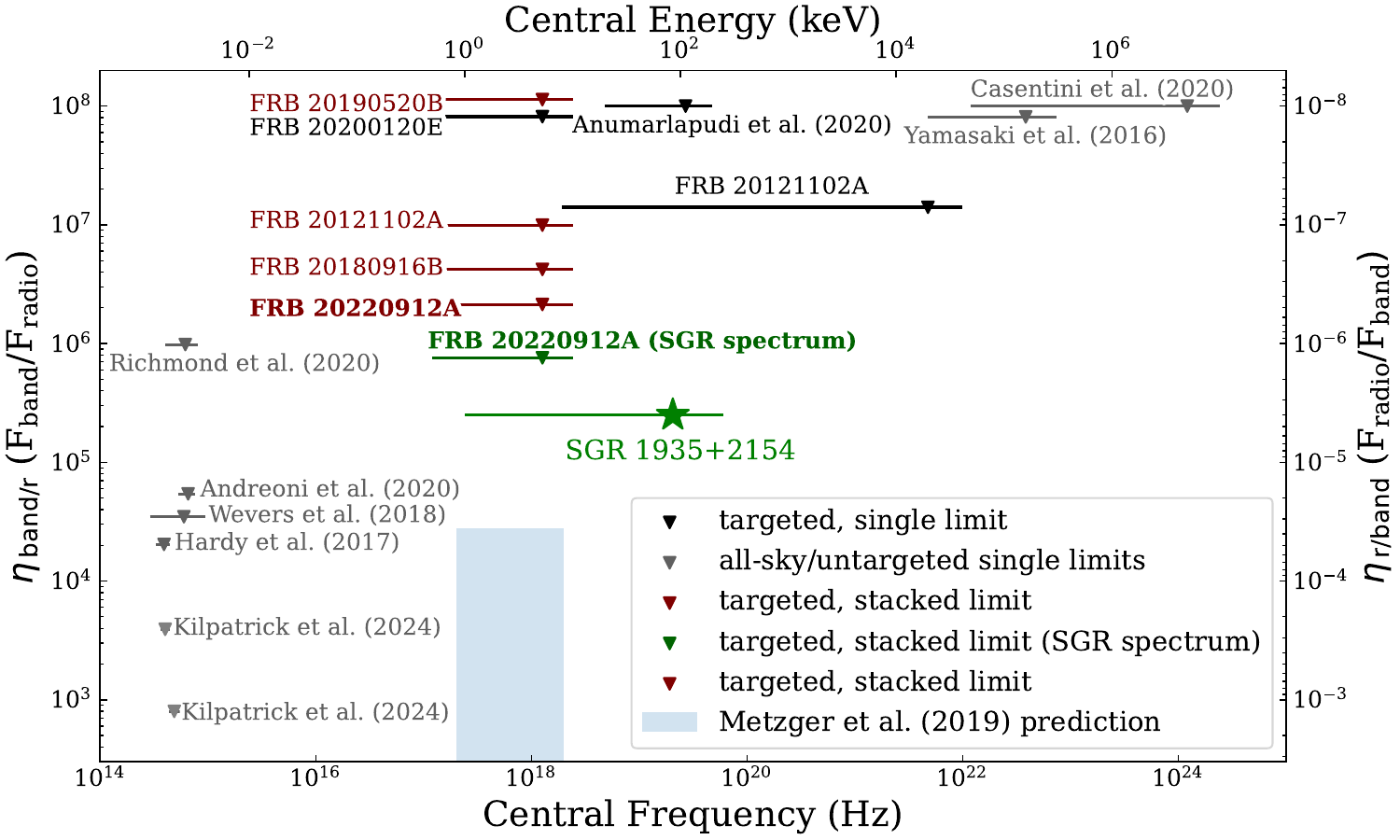}
    \caption{Upper limits placed on the band-integrated HE-to-radio fluence $(F)$ ratio ($\eta_{\text{band/r}} = F_{\text{band}}/F_{\text{Radio}}$, since we also consider measurements outside of the X-ray band) for optical, UV, X-ray and gamma-ray observations during simultaneous radio observations of FRBs, adapted from \cite{2020ApJ...897..146C}. For \repeater\,, the limit presented is calculated using the Bayesian method described in Appendix \ref{app:eta}, considering all 30 upper limits at the time of radio bursts and assuming a 10 keV blackbody burst spectrum. For consistency, we present upper limits from previously reported non-detections combined using the Bayesian stacking method presented in this paper (maroon lines, bold text to indicate the values derived from our observations).  Other relevant X-ray and gamma-ray limits, based on targeted observations, are shown as black lines
     \citep{r1limits, 2020ApJ...888...40A, 2023arXiv230810930P}.
    Previous constraints from untargeted observations are labeled by publication in gray \citep{2020MNRAS.491.5852A,  2020ApJ...890L..32C,
    2017MNRAS.472.2800H,
    2020PASJ...72....3R,
    r1limits,
    2020ApJ...901..165S,
    2018MNRAS.473.3854W,
    2016MNRAS.460.2875Y}. We show the measured $\ate$ of the X-ray burst observed contemporaneously with FRB-like emission from Galactic magnetar SGR 1935+2154 detected on 2020 April 28 (green star; \citealt{2021NatAs...5..378L}) and the forest green line shows our stacked $\ate$ limit assuming the observed burst spectrum of this SGR 1935+2154 X-ray burst \citep{2021NatAs...5..378L}.
    The predicted parameter space of $\ate$ for a $\sim$keV-energy young magnetar burst\,\citep{2019MNRAS.485.4091M}, assuming a luminosity distance of 362.4\,Mpc for \repeater\, and the observed radio fluence of radio burst B30, is shown in blue.}
    \label{fig:phasespace}
\end{figure*}

For all 30 radio bursts from \repeater\, that were detected during our X-ray observations, we compute this stacked upper limit of $2 \times 10^6$ on $\ate$ at the 99.7\% credible level, assuming our conservative 10 keV blackbody burst spectrum. The posterior distribution of this model is derived in full in Appendix \ref{app:eta}. The observed SGR 1935-2154 X-ray burst associated with FRB-like emission was modelled with a cutoff powerlaw at $E_{\text{cut}}=83.89$ keV, photon index $\Gamma = 1.56$ \citep{2021NatAs...5..378L}. If we instead assume this spectrum for our burst model, and after having corrected for absorption,  we derive an upper limit on $\ate$ of ${8 \times 10^5}$ at the 99.7\% credible level. We show the posterior \anedit{distributions} on $\ate$ using this method in the top right panel of Figure \ref{fig:posteriors} and compare with previous `\textit{bona fide}' counterpart limits in Figure \ref{fig:phasespace}. For the 2004 December 27 magnetar giant flare of SGR 1806-20, \cite{2016ApJ...827...59T} placed upper limits on the possible radio fluence at the time of the 1.4 erg cm$^{-2}$ X-ray flare given the non-detection with the Parkes radio telescope, which was observing a location 35.6 degrees away from the source at the time \citep{2005Natur.434.1098H,2005Natur.434.1107P,2005Natur.434.1110T}. This corresponds to a \textit{lower}-limit on X-ray-to-radio fluence ratio, given it is an upper limit on the radio fluence which is the denominator, of $4 \times 10^{10}$ for X-ray counterparts of FRBs if they can be attributed to magnetar giant flares like that observed for SGR 1806-20. However, the majority of upper limits placed in that paper are incompatible with our limits and all published limits on $\ate$ to date. 

Using the formalism derived in Appendix \ref{app:eta}, we revisit the limits placed by \cite{r1limits, Pilia_2020,  2020ApJ...901..165S}, and \cite{2024arXiv240212084Y}. 
This allows a more direct comparison between the limits, compiles all the data in one place, and decreases the upper limit placed on $\ate$ for each work. For the aforementioned papers, Table \ref{tab:paulaaron} summarizes the data that are required to calculate our stacked $\ate$ constraint (Equation \ref{eqn:etalikelihood}). The full posterior \anedit{distribution}s from the retreatment of these soft-X-ray data are shown in Figure \ref{fig:posteriors}. The corresponding 99.7\% credible intervals on $\ate$ from these data, using our method, are also placed in the broader context of observing frequency-$\ate$ phase space in Figure \ref{fig:phasespace}. 

The stacked $\ate$ limit placed on \rthree\, plotted in Figure \ref{fig:phasespace} uses the observations of \cite{r1limits} and \cite{Pilia_2020}. The limit could be improved by the inclusion of limits placed by \cite{2023AnA...676A..17T}, but the required data are not currently public. 

\section{Discussion}
\label{sec:discussion}

We now discuss each of the limits derived from our observations in the context of known transient populations and predictions from FRB models. \cite{2019MNRAS.485.4091M} predict an 1--10 keV X-ray counterpart with luminosity $10^{42}$--$10^{43}$ erg s$^{-1}$ for their model of FRBs as synchrotron maser emission from decelerating relativistic blast waves. When one places this emission, predicted to last 0.1--1 seconds, at the distance of \repeater\,, it corresponds to fluences in the range $(6-600)\times 10^{-15}$ erg cm$^{-2}$. 

Our most constraining 99.7\% credible region upper limit on X-ray emission at the time of a radio burst from FRB 20220912A, $1.5 \times 10^{-10}$ erg cm$^{-2}$, does not yet probe this region. However, our simultaneous X-ray and radio measurements of the hyperactive, bright FRB 20220912A allow us to place the best $\ate$ constraints in the X-ray band to date. Our lowest 99.7\% $\ate$ upper limit is ${7 \times 10^{6}}$, and when stacking data from the time of each of our radio bursts, is $2\times10^6$. While the limits in this paper are not the most constraining in X-ray burst luminosity placed for an FRB to date, owing to the larger distance of \repeater\, compared to \meor\ \citep{2023arXiv230810930P}, our limits remain highly relevant because the hyperactivity and brightness of the source allow us to place deeper limits on $\ate$.

Our observations cannot rule out magnetospheric models, which, if they predict an X-ray counterpart, cite expected $\ate$ from $\sim$ 1 (comparable energy to that of the radio burst, e.g., \citealt{2020ApJ...899L..27M}) to $\ate \sim 10^{4}$ \citep{10.1093/mnras/staa2450}. Drawing on analogies with solar flares, \cite{2002ApJ...580L..65L} hypothesized that bursts from magnetars could produce an X-ray-to-radio fluence ratio of $\sim 10^{4}$ (this model was later employed as an explanation of FRBs by \citealt{2010vaoa.conf..129P}). This analogy was recently further contextualized for the microshots emitted by this source, FRB 20220912A, by \cite{10.1093/mnras/stad2847}. 
Our best $\ate$ limit is the closest yet to the $\ate = 2.5\times10^5$  observed for the FRB-like burst on 2020 April 28 from SGR\,1935+2154 and accompanying X-ray burst \citep{2020Natur.587...54C, 2020Natur.587...59B, Mereghetti_2020, 2021NatAs...5..378L}. Indeed, for a similar X-ray burst spectrum (a cutoff powerlaw at $E_{\text{cut}}=83.89$ keV, photon index $\Gamma = 1.56$, and after having corrected for galactic absorption; \citealt{2021NatAs...5..378L}), instead of our conservative 10 keV blackbody burst spectrum, the stacked $\ate$ limit is $\mathbf{8} \times 10^5$, only a factor of \anedit{3} from the observed $\ate$ of the FRB-like burst from SGR\,1935+2154. This motivates the continued search for X-ray counterparts for FRB sources. After a statistically significant number of bright radio bursts like those detected from \repeater, one could disfavor the mechanism producing the SGR 1935+2154 simultaneous X-ray and radio burst for a given repeater source if no X-ray emission was seen, under the assumption that an X-ray burst with equal $\ate$ is emitted with each radio burst. Considering simultaneous \xmm \, and \eff\, observations like the campaign in this paper, $\ate\sim10^4$ could be measured with our Bayesian method at the 99.7\% level given an X-ray non-detection of two kJy ms radio bursts, ten 500 Jy ms radio bursts, or fifty 50 Jy ms radio bursts. Highly energetic bursts from repeaters are detected more rarely, but they have been observed before \citep{2024NatAs...8..337K}.\cite{10.1093/mnras/stad2847} and \cite{ 2022ATel15817....1O} report a handful of bursts from \repeater\, with radio fluences $> 400$ Jy ms and as high as 972 Jy ms. \cite{10.1093/pasj/psac101} report a burst from FRB 20201124A with fluence $> 189$ Jy ms. 

In 2020 October, SGR 1935+2154 was observed to emit regular pulsed radio emission, with radio bursts detected with luminosities comparable to typical rotating radio transients or radio pulsars, depending on which distance was assumed \citep{2023SciA....9F6198Z}. This pulsed radio emission was anti-aligned with the X-ray pulsed emission at the time \citep{2023NatAs...7..339Y}. 
We computed the fraction of X-ray-to-radio energy released during an average single-pulse, and find this $\ate$ of $\sim (7-12) \times 10^6$, which can be ruled out for \repeater\, by our observations \citep{2023SciA....9F6198Z,2023NatAs...7..339Y}.

Given that \repeater\, and \rthree\, have source distances of the same order of magnitude, and were observed with the same telescopes at the times of radio bursts, the upper limit we derive on persistent X-ray emission at the location of \repeater\, is similar in magnitude to that derived for \rthree\, \citep{2020ApJ...901..165S}. Thus we can derive similar conclusions about the nature of the source. A Crab-like nebula, which has persistent X-ray luminosity $\sim 10^{37}$ erg s$^{-1}$, cannot be ruled out for the source. Our limit is lower than the luminosities of most ULXs, but we cannot rule out luminosities in the range of Galactic HMXBs and LMXBs \citep{2003ApJ...583..145T, 2018MNRAS.476.2530S, 10.1093/mnras/sty3403, 2023arXiv230810930P}.

\section{Conclusions}

High energy studies of FRB counterparts are crucial to derive a full picture of the spectral properties of FRBs, and hence not only to disentangling the sources of FRBs but also as a probe of one the most extreme radio transients in the Universe. CHIME/FRB is a unique monitor of stochastic repeater activity, which is often clustered in time \citep[e.g.,][]{2020Natur.582..351C, 2022ApJ...927...59L,2024ATel16420....1S}. Such a monitor allows coordination of HE observations with the maximum probability of detecting contemporaneous radio bursts--- this, along with the Bayesian stacking methodology presented in this paper enables searches in new areas of counterpart relative-fluence ($\ate$) phase space.

The recently hyper-activated \repeater\, is an example that shows the power of these types of observations. Based on an extensive, contemporaneous radio and X-ray campaign, we report our lowest single 99.7\% credible upper limit on $\ate$ of ${7\times10^6}$, which is the lowest constraint yet.  Using a hierarchical extension to the standard Bayesian treatment of low-count, background contaminated data, we combined information from X-ray non-detections at the times of 30 of radio bursts from FRB 20220912A. This allowed us to constrain $\ate < 2\times10^6$ at the 99.7\% level, assuming all bursts are associated with X-ray bursts with the same fluence ratio. Our brightest radio burst observed simultaneously with an X-ray telescope produces a 99.7\% credible region 0.5--10.0 keV fluence limit of \anedit{${1.5\times10^{-10}}$ erg cm$^{-2}$} assuming a 100ms-burst with a 10 keV blackbody spectrum, corrected for absorption by a ${10^{22}}$ cm$^{-2}$ neutral hydrogen column. Our \xmm\, observations constrain, at the 99.7\% level, persistent flux from \repeater\, to less than \anedit{${8.8\times 10^{-15}}$ ergs cm$^{-2}$ s$^{-1}$} in the 0.5--10-keV range assuming \anedit{a powerlaw spectrum with $\Gamma = 2$ after correcting for absorption by a ${10^{22}}$ cm$^{-2}$ neutral hydrogen column}.  At the luminosity distance of 362.4 Mpc, this corresponds to a 0.5--10.0 keV luminosity of $L_X < {1.4 \times 10^{41}}$ ergs s$^{-1}$.

Continued arcsecond localizations from projects like the Deep Synoptic Array-110~(DSA-110) and the upcoming CHIME/FRB Outriggers will allow us to target the most active, nearby, and energetic of FRB sources for HE follow-up campaigns like the one detailed in this paper (\citealt{2023arXiv231010018B, 2023arXiv230703344L, 2024arXiv240207898L}). This will allow the ongoing pursuit of source-discriminating HE emission closer to the luminosities predicted by many models and observed transient behavior from magnetars in our own Galaxy (e.g., see~\citealt{2023arXiv230810930P}).

\section*{Acknowledgments}
We express our sincere gratitude to the \eff, \nicer, \swift, and \xmm\ operations teams for their help coordinating these observations and their remarkable response times. We are deeply grateful to Keith Gendreau, Zaven Arzoumanian, and Elizabeth Ferrera for promptly scheduling these \nicer\ observations and for their support of our work. We thank Alex Kraus for helping to schedule these Effelsberg observations. We also thank Aaron Tohuvavohu for numerous useful discussions and Ziggy Pleunis for helpful comments, both of which improved the quality of the manuscript. 

A.M.C. is funded by an NSERC Doctoral Postgraduate Scholarship. 
A.B.P. is a Banting Fellow, a McGill Space Institute~(MSI) Fellow, and a Fonds de Recherche du Quebec -- Nature et Technologies~(FRQNT) postdoctoral fellow.
The Dunlap Institute is funded through an endowment established by the David Dunlap family and the University of Toronto. B.M.G. acknowledges the support of the Natural Sciences and Engineering Research Council of Canada (NSERC) through grant RGPIN-2022-03163, and of the Canada Research Chairs program.
F.A.D is supported by the UBC Four Year Fellowship.
G.M.E. acknowledges funding from NSERC through
Discovery Grant RGPIN-2020-04554.
V.M.K. holds the Lorne Trottier Chair in Astrophysics \& Cosmology, a Distinguished James McGill Professorship, and receives support from an NSERC Discovery grant (RGPIN 228738-13), from an R. Howard Webster Foundation Fellowship from CIFAR, and from the FRQNT CRAQ.
FRB research at UBC is supported by an NSERC Discovery Grant and by the Canadian Institute for Advanced Research.
M.B. is a McWilliams fellow, an International Astronomical Union Gruber fellow, and receives support from the McWilliams seed grant.
A.P.C is a Vanier Canada Graduate Scholar.
K.W.M. holds the Adam J. Burgasser Chair in Astrophysics and is supported by NSF grants (2008031, 2018490).
A.P. is funded by the NSERC Canada Graduate Scholarships -- Doctoral program.
K.S. is supported by the NSF Graduate Research Fellowship Program.
M.W.S. acknowledges support from the Trottier Space Institute fellowship program.
D.C.S. is supported by an NSERC Discovery Grant (RGPIN-2021-03985). B.M.G., D.C.S., G.M.E. acknowledge additional support provided by the Canadian Statistical Sciences Institute through the funding of an interdisciplinary Collaborative Research Team.

This publication is partly based on observations with the 100-m telescope of the MPIfR (Max-Planck-Institut f\"ur Radioastronomie) at Effelsberg. This work made use of data supplied by the UK Swift Science Data Centre at the University of Leicester. This work was partly based on observations obtained with \xmm, an ESA science mission with instruments and contributions directly funded by ESA Member States and NASA.

We acknowledge that CHIME is located on the traditional, ancestral, and unceded territory of the Syilx/Okanagan people. We are grateful to the staff of the Dominion Radio Astrophysical Observatory, which is operated by the National Research Council of Canada.  CHIME is funded by a grant from the Canada Foundation for Innovation (CFI) 2012 Leading Edge Fund (Project 31170) and by contributions from the provinces of British Columbia, Qu\'{e}bec and Ontario. The CHIME/FRB Project is funded by a grant from the CFI 2015 Innovation Fund (Project 33213) and by contributions from the provinces of British Columbia and Qu\'{e}bec, and by the Dunlap Institute for Astronomy and Astrophysics at the University of Toronto. Additional support was provided by the Canadian Institute for Advanced Research (CIFAR), McGill University and the Trottier Space Institute thanks to the Trottier Family Foundation, and the University of British Columbia.

\software{pwkit \citep{2017ascl.soft04001W}, astropy \citep{2022ApJ...935..167A}}
\facilities{CHIME, Effelsberg, \textit{NICER}, \textit{Swift}, and \textit{XMM-Newton}}

\begin{deluxetable*}{ccc}
\caption{\label{tab:params} Source parameters used in the analysis}
\tablehead{
\colhead{Parameter} & \colhead{Value} & \colhead{Reference}}
\startdata
TNS Name & FRB\,20220912A & \cite{2022ATel15679....1M,2021AAS...23742305G}\tablenotemark{a}\\
R.A. (J2000)\tablenotemark{b}& 23$^{\text{h}}$09$^{\text{m}}$04.8988(50)$^{\text{s}}$ & \cite{2023arXiv231214490H}  \\
Decl. (J2000)\tablenotemark{b} & 48$^\circ$42'23.9078(50)'' &\cite{2023arXiv231214490H}  \\
DM & 219.456(5) \dmunits & \cite{2022ATel15679....1M} \\
Redshift & 0.0771(1) &  \cite{2023ApJ...949L...3R} \\
Luminosity distance & 362.4(1) Mpc& \cite{2023ApJ...949L...3R} 
\enddata
\tablenotetext{a}{https://www.wis-tns.org/}
 \tablenotetext{b}{While the positions reported by \cite{2023ApJ...949L...3R} and \cite{2023arXiv231214490H} agree on the host galaxy of the repeater, they are not consistent, within their errors, with one another. We assume the position reported by the European VLBI Network \citep{2023arXiv231214490H} throughout this paper, but have performed the analysis with both positions and find no X-ray detection in either case.}
\end{deluxetable*}
\bibliography{xraybib}{}
\bibliographystyle{aasjournal}
\appendix
\section{Bayesian formalism for source rate constraints from multiple observations} 
\label{app:rate}
We seek to combine information from $n$ X-ray observations at the time of radio bursts, which can be treated as independent trials if sufficiently separated in time (i.e., the time between the FRBs is long compared to the durations tested). We follow the \cite{kbn} Bayesian formalism, but construct the posterior \anedit{distribution} probability using Bayes rule assuming $n$ observations of $N_i$ X-ray photons for $(i \in 1, \ldots n)$, and (known) average background rates of $B_i$ for $(i \in 1, \ldots, n)$, to estimate a rate $S$ for X-ray emission at the time of radio bursts. This method assumes that $S$ is constant for all radio bursts, during the radio bursts themselves. From Bayes rule, we can derive the posterior \anedit{distribution} $f$ on the rate $S$:

\begin{align}
p(S) & \propto 1\\
N_i & \sim \text{Poisson}(\lambda_i = B_i + S)\\
    f(S| N_1, \ldots, N_{n}) & =\frac{ P( N_1, \ldots, N_{n}|S)p(S)  }{P(N_1 \ldots, N_n)}\\
    & \propto \prod_{i=1}^{n} \text{Poisson}(N_i | \lambda_i = B_i + S), 
\end{align}
where $\text{Poisson}(k| \lambda)$ is the Poisson probability mass function of $k \in \mathbb{N}$ observed counts given Poisson rate $\lambda \in \mathbb{R}^+$ and $B_i$ is assumed constant and known for each observation $n$. \cite{kbn} assume an improper uniform prior \anedit{distribution} p(S) = c for all positive values. In order to construct a proper (finite) uniform prior distribution, one should instead assume $p(S) = c \ \forall S \in [0,x]$ for some appropriately large value $x \in \mathbb{R}^+$ of X-ray rate, and practically, this will be enforced by any analytic implementation which computes this posterior \anedit{distribution}. 
Previous limits placed on the X-ray emission at the time of FRBs can instead be used to set a conservative but still informative prior \anedit{distribution}. 
In order to construct the 99.7\% credible interval from this equation, again following \cite{kbn}, we enforce that the difference in the flux bounds of the credible interval ($S_{\text{max}}-S_{\text{min}}$) is minimized and the peak value of the posterior \anedit{density distribution} is included. This is known as the highest posterior density interval \citep[see, e.g., chapter one of][for applications in astronomy]{2018sabm.book...29S}. 
\section{Bayesian formalism for $\ate$ constraints from multiple observations} 

\label{app:eta}

In Appendix \ref{app:rate}, we assume a single source rate, $S$, and calculate the posterior \anedit{distribution} on that source rate by combining information from multiple observations. If one expects that the X-ray fluence from the source should be proportional to the radio fluence, it is desirable to estimate the posterior \anedit{distribution} on the relative X-ray to radio source fluence, $\ate$. We thus assume here that $\ate$ is constant for all bursts, and that $S_i$ can vary. Again, we define $N_i$ as the number of X-ray photons at the time of each radio burst, $B_i$ is the average background rate of X-ray photons at the time of each radio burst and $F_{\text{radio, }i}$ is the calculated radio fluence, with associated error $\sigma_{F_{\text{radio, }i}}$. What is the credible interval on $\ate$ for these multiple observations? In the following derivation, we use $N_i, B_i, F_i$ as shorthand for the more conventional general list $N_0, B_0, F_0, \ldots, N_n, B_n, F_n$ where $n$ is the total number of detected bursts. We will compute the posterior \anedit{distribution} of the observations given following model, introduced in Section \ref{sec:stacked_section}:

\begin{align}
  \text{Level I} &&  N_i & \sim \text{Poisson}(\lambda_i = S_i + B_i)\\
   \text{Level II} && S_i & \sim \mathcal{N}^\infty_0  \left( \frac{\ate F_{\text{radio}, i}}{(\text{Flux/S})} , \frac{\ate \sigma_{F_{\text{radio}, i}}}{(\text{Flux/S})} \right), 
\end{align}
where (Flux/S)$\in \mathbb{R}^+$ is a conversion parameter to turn the X-ray count rates into fluxes. This value depends on the underlying spectral model assumed and the effective area of the X-ray telescope, but can be computed using standard X-ray tools. We assume a blackbody spectral model with $kT = 10$\,keV for the bursts and present the corresponding (Flux/S) parameter for each observation in Table \ref{tab:paulaaron}.  $\mathcal{N}_0^\infty (\mu, \sigma)$ denotes a normal distribution truncated on the left at $0$ with mean $\mu \in \mathbb{R}$ and standard deviation $\sigma \in \mathbb{R}^+$. The truncation is introduced because negative source counts are not physical. Poisson$(\lambda)$ denotes the Poisson distribution with rate parameter $\lambda \in \mathbb{R}^+$. The $B_i$ are treated as known and fixed, but an additional model for the error can be added in Level II if there are significant uncertainties in this estimation or the background rate is variable. 
Starting again from Bayes rule, we can write the posterior \anedit{distribution} \anedit{distribution} $f(\ate)$

\begin{align}
    f(\ate) & = \frac{p(N_i|\ate)p(\ate)}{C},
\end{align}
 where $C\in \mathbb{R}$ is some normalization constant. To compute the probability density of the observed counts, we must invoke an X-ray rate parameter for each observation, $S_i$, however this value is not known. Instead, we assume a hierarchical Bayesian model, introducing $S_i$ as a random variable in the following equation using the chain rule of probability through the identity $p(A|B) = \int_C p(A,C|B) \dd C = \int_C p(C|B) p(A|B,C)\dd C$ for random variables $A,B,C$:

\begin{align}
 f(\ate) & = \frac{1}{C} p(\ate) \int_{S_1}\int_{S_2} \ldots \int_{S_n} p(N_i | S_i, \ate) p(S_i | \ate) \dd S_1\dd S_2\ldots \dd S_n\\
    & \propto p(\ate) \int_{S_1}\int_{S_2} \ldots \int_{S_n} p(N_i | S_i,\ate) p(S_i | \ate) \dd S_1\dd S_2\ldots \dd S_n\\
    & \propto p(\ate)\int_{S_1}\int_{S_2} \ldots \int_{S_n} \prod_i \text{Pois}(N_i | \lambda_i = S_i+B_i) \hspace{1.5mm} \mathcal{N}^\infty_0  \left( \frac{\ate F_{\text{radio}, i}}{(\text{Flux/S})} , \frac{\ate \sigma_{F_{\text{radio}, i}}}{(\text{Flux/S})} \right)\dd S_1\dd S_2\ldots \dd S_n\\
    & \propto p(\ate) \prod_i \int^\infty_{0} 
      \hspace{1.5mm} 
      \frac{(\text{Flux/S})(B_i + S_i)^{N_i}}{ \sqrt{2\pi}\ate \sigma_{F_{\text{radio}, i}} (N_i!)} \exp[-\left(B_i + S_i +\frac{(\text{Flux/S})^2\left(S_i - \frac{\ate F_{\text{radio}, i}}{(\text{Flux/S})}\right)^2}{2\ate^2 \sigma^2_{F_{\text{radio}, i}}}\right)]
 \dd S_i\\
     & \propto p(\ate) \exp(-\sum_{i=1}^n  B_i) \prod_{i=1}^{n} \int^\infty_{0} 
      \hspace{1.5mm} 
      \frac{(\text{Flux/S})(B_i + S_i)^{N_i}}{ \sqrt{2\pi}\ate \sigma_{F_{\text{radio}, i}} (N_i!)} \exp[-\left( S_i +\frac{(\text{Flux/S})^2\left(S_i - \frac{\ate F_{\text{radio}, i}}{(\text{Flux/S})}\right)^2}{2\ate^2 \sigma^2_{F_{\text{radio}, i}}}\right)]
 \dd S_i\\
 & \propto p(\ate)  \prod_{i=1}^{n} \int^\infty_{0} 
      \hspace{1.5mm} 
      \frac{(B_i + S_i)^{N_i}}{\ate } \exp[-\left( S_i +\frac{(\text{Flux/S})^2\left(S_i - \frac{\ate F_{\text{radio}, i}}{(\text{Flux/S})}\right)^2}{2\ate^2 \sigma^2_{F_{\text{radio}, i}}}\right)]
 \dd S_i. \label{eqn:etalikelihood}
\end{align}
This expression can be numerically integrated directly and normalized, or estimated with MCMC methods. We use the posterior \anedit{distribution} from a previous independent trial as our prior \anedit{distribution} when stacking. 

The reported credible regions correspond to the highest posterior density interval. A numerical implementation of this integral in python and a minimal working example \anedit{is available on Zenodo at} \url{https://doi.org/10.5281/zenodo.12785591},  \citep{cook202412785591}

\begin{longrotatetable}
\begin{deluxetable*}{llcllllllll}
\tablecaption{\label{tab:paulaaron} Summary of observations from previous X-ray counterpart searches of FRBs. All values are reported for an X-ray observing band $0.5-10.0$keV. In all studies, there are no X-ray photons detected coincident with the source position within 100 ms of each burst. The corresponding stacked $
\ate$ credible interval upper limits can be found in Figure \ref{fig:phasespace}. All arrival times are barycentric and corrected to infinite frequency. We assume an error of 10\% on the fluences of \cite{r1limits}. \cite{2024arXiv240212084Y} report only the radio burst energy, and hence we estimate the radio burst fluence assuming the full bandwidth of FAST, 400 MHz. The multiwavelength campaigns of \cite{2021AnA...656L..15P} and \cite{2023AnA...676A..17T} are also appropriate for direct comparison here, however the summary statistics in these papers are not enough to perform the Bayesian analysis and the data are not publicly available.}
\tablehead{\colhead{Source \hspace{8mm} Publication}  & \colhead{Arrival time} &  \colhead{background rate} & \colhead{X-ray telescope} & \colhead{Radio fluence}& \colhead{count to fluence} & \colhead{Bandwidth} &\colhead{$\eta_{\text{\,x/r, single}}$\tablenotemark{a}}&
\colhead{$\eta_{\text{\,x/r, stack}}$\tablenotemark{a}}\\
 \colhead{} & \colhead{MJD}  & \colhead{counts/s} & \colhead{} & \colhead{Jy ms}& \colhead{erg cm$^{-2}$/count} & \colhead{MHz} }
\startdata
FRB 20121102A \hspace{6mm}  &  &  &   &    &  & && $1\times10^7$\\
\hspace{12mm} \cite{r1limits} &  &  &   &    &    \\ 
 & 57647.232346450619 & $1 \times 10^{-2}$ &\xmm& $0.82(8)$ & $1.4\times 10^{-11}$& 600 & $2 \times 10^7$& \\ 
& 57647.232346883015 & $1 \times 10^{-2}$ &\xmm& $0.16(2)$ & $1.4\times 10^{-11}$ & 600 & $9 \times 10^7$& \\ 
& 57649.173812898174 & $5 \times 10^{-2}$ &\xmm&1.3(1) & $1.4\times 10^{-11}$ & 600 & $1 \times 10^7$& \\ 
& 57649.218213226581 & $5 \times 10^{-2}$ &\xmm & 0.34(3) & $1.4\times 10^{-11}$ & 600 & $4 \times 10^7$& \\ 
& 57765.049526345771 &  $5 \times 10^{-5}$ &\textit{Chandra} & 0.33(3) & $3.5\times 10^{-11}$ & 600 & $1 \times 10^8$& \\ 
& 57765.064793212950 &  $5 \times 10^{-5}$ &\textit{Chandra} & 0.83(8) & $3.5\times 10^{-11}$ & 600 &  $4 \times 10^7$&\\ 
& 57765.069047502300 &  $5 \times 10^{-5}$ &\textit{Chandra}&  0.62(6) &$3.5\times 10^{-11}$ & 600 & $5 \times 10^7$ & \\ 
& 57765.120778204779 &  $5 \times 10^{-5}$ &\textit{Chandra} &  1.1(1) & $3.5\times 10^{-11}$ & 600 & $3 \times 10^7$ & \\ 
& 57765.136498608757 &  $5 \times 10^{-5}$ &\textit{Chandra} &  0.22(2) & $3.5\times 10^{-11}$ & 600 & $2 \times 10^8$& \\ 
& 57765.100827849608 &  $5 \times 10^{-5}$ &\textit{Chandra} & 0.37(4) & $3.5\times 10^{-11}$& 600 &$1 \times 10^8$ & \\ 
& 57765.108680842022 &  $5 \times 10^{-5}$ &\textit{Chandra} & 0.030(3) & $3.5\times 10^{-11}$& 600 & $2 \times 10^9$& \\ 
& 57765.143337535257 &  $5 \times 10^{-5}$ &\textit{Chandra} & 0.10(1) & $3.5\times 10^{-11}$ & 600 &$3 \times 10^8$ & \\ 
FRB 20180916B & &  &  &   &    &  & & $4\times10^6$ \\
\hspace{12mm} \cite{2020ApJ...901..165S} &  &  &  &  &  &  & & \\
 &  58835.17721035 & $ 6 \times 10^{-5}$ &\textit{Chandra}& $2.9(7)$ & $6.3 \times 10^{-11}$ & 200 & $2 \times 10^8$ \\
  \hspace{12mm} \cite{Pilia_2020} &  &  &  &  &  &  &  \\
 &58899.56141184  &$1\times10^{-3}$ & \xmm & $37(16)$ &  $2.2 \times 10^{-11}$ &  60  &   $7\times10^6$ \\ 
  &58899.56781756 & $6\times10^{-3}$& \xmm  &  $13(8)$& $2.2 \times 10^{-11}$  & 60   &   $3 \times 10^7$ \\ 
&58899.57561573 & $8\times10^{-3}$& \xmm &   $19(8)$ & $2.2 \times 10^{-11}$ &  60  & $1\times10^7$ &  \\ 
FRB 20190520B \hspace{6mm}  &  &  &   &    &  & && $1\times10^8$\\
\hspace{12mm} \cite{2024arXiv240212084Y} &  &  &   &    &    \\
& 58991.7353560987 & $5 \times 10^{-3}$ &\swift& 0.237(1) & $1.94\times 10^{-10}$& 400 & $1 \times 10^9$& \\ 
& 59067.4866461265 & $5 \times 10^{-3}$ &\swift& 0.224(1) & $1.89\times 10^{-10}$ & 400 & $1 \times 10^9$& \\ 
& 59067.4866467051 & $5 \times 10^{-3}$ &\swift& 0.251(1) & $1.89\times 10^{-10}$& 400 & $1 \times 10^9$& \\ 
& 59071.4724304051 & $1.5 \times 10^{-2}$ &\swift& 0.135(1) & $1.93\times 10^{-10}$ & 400 & $2 \times 10^9$& \\ 
& 59071.4724306366 & $1.5 \times 10^{-2}$ &\swift& 0.114(1) & $1.93\times 10^{-10}$& 400 & $2 \times 10^9$& \\ 
& 59075.4542594866 & $2 \times 10^{-2}$ &\swift& 0.535(1) & $1.89\times 10^{-10}$ & 400 & $5 \times 10^8$& \\ 
& 59075.4547689533 & $2.5 \times 10^{-2}$ &\swift& 0.359(1) & $1.96\times 10^{-10}$& 400 & $7 \times 10^8$& \\ 
& 59077.4977122303 & $2 \times 10^{-2}$ &\swift& 0.237(1) & $1.96\times 10^{-10}$ & 400 & $1 \times 10^9$& \\ 
& 59077.4978634245 & $2 \times 10^{-2}$ &\swift& 0.154(1) & $1.96\times 10^{-10}$& 400 & $2 \times 10^9$& \\ 
& 59077.4988665859 & $2.5 \times 10^{-2}$ &\swift& 0.161(1) & $1.96\times 10^{-10}$ & 400 & $2 \times 10^9$& \\ 
\repeater & &  &  &   &    &  & &$2 \times 10^6$ \\
\hspace{12mm} this work &  &  &  &    &  &  &  \\
&   59867.2337651529(5) & $1.5 \times 10^{-2}$ & \swift & $>1.12$ & ${2.0 \times 10^{-10}}$& 262 & ${3 \times 10^8}$\\ 
&   59867.2361652994(5) & $2.0 \times 10^{-2}$ & \swift & $>0.52$ & ${2.0 \times 10^{-10}}$& 262 & ${6 \times 10^8}$\\ 
&   59868.2302000210(5) & $2.0 \times 10^{-2}$ & \swift & $>0.24$ & ${2.1 \times 10^{-10}}$& 262 & ${1 \times 10^9}$ \\ 
& 59880.1990692887(5) & $3.0\times 10^{-2}$ & \nicer & $> 1.27$ & ${4.4 \times 10^{-11}}$& 262 &${5\times 10^{7}}$ \\ 
& 59880.2007281066(5) & $2.6\times 10^{-2}$ & \nicer & $> 0.32$ & ${4.4 \times 10^{-11}}$& 262 &${2\times 10^{8}}$ \\ 
& 59880.2012657886(5) & $2.8\times 10^{-2}$ & \nicer & $> 0.45$ & ${4.4 \times 10^{-11}}$& 262 &${1\times 10^{8}}$ \\ 
& 59880.2014033844(5) & $2.8\times 10^{-2}$ & \nicer & $> 0.42$ & ${4.4 \times 10^{-11}}$& 262 &${2\times 10^{8}}$ \\ 
& 59880.2021430438(5) & $3.2\times 10^{-2}$ & \nicer & $> 0.42$ & ${4.4 \times 10^{-11}}$& 262 &${2\times 10^{8}}$ \\ 
& 59880.2021439275(5) & $3.2\times 10^{-2}$ & \nicer & $> 0.56$ & ${4.4 \times 10^{-11}}$& 262 &${1\times 10^{8}}$ \\ 
& 59880.2039849901(5) & $2.8\times 10^{-2}$ & \nicer & $> 0.91$ & ${4.4 \times 10^{-11}}$& 262 &${7\times 10^{7}}$ \\ 
& 59880.2045395375(5) & $4.2\times 10^{-2}$ & \nicer & $> 1.13$ & ${4.4 \times 10^{-11}}$& 262 &${6\times 10^{7}}$ \\ 
& 59882.1909163952(5) & $2.2\times 10^{-2}$ & \nicer & $> 1.97$ & ${4.4 \times 10^{-11}}$& 262 &${3\times 10^{7}}$ \\
& 59882.1915613202(5) & $4.2\times 10^{-2}$ & \nicer & $> 0.67$ & ${4.4 \times 10^{-11}}$& 262 &${1\times 10^{8}}$ \\ 
& 59882.1951750479(5) & $3.8\times 10^{-2}$ & \nicer & $> 0.66$ & ${4.4 \times 10^{-11}}$& 262 &${1\times 10^{8}}$ \\
& 59882.1951762690(5) & $3.8\times 10^{-2}$ & \nicer & $> 0.70$ & ${4.4 \times 10^{-11}}$& 262 &${1\times 10^{8}}$ \\ 
& 59882.1951801563(5) & $3.8\times 10^{-2}$ & \nicer & $> 0.83$ & ${4.4 \times 10^{-11}}$& 262 &${8\times 10^{7}}$ \\ 
& 59882.1951805052(5) & $3.8\times 10^{-2}$ & \nicer & $> 0.43$ & ${4.4 \times 10^{-11}}$& 262 &${2\times 10^{8}}$ \\ 
& 59882.1955737831(5) & $3.0\times 10^{-2}$ & \nicer & $> 0.79$ & ${4.4 \times 10^{-11}}$& 262 &${8\times 10^{7}}$ \\
& 59884.1880391532(5) & $1.4\times 10^{-2}$ & \nicer & $> 0.78$ & ${4.4 \times 10^{-11}}$& 262 &${8\times 10^{7}}$ \\ 
& 59884.1902762081(5) & $2.0\times 10^{-2}$ & \nicer & $> 0.36$ & ${4.4 \times 10^{-11}}$& 262 &${2\times 10^{8}}$ \\ 
& 59884.1908769493(5) & $1.4\times 10^{-2}$ & \nicer & $12.3(14)$ & ${4.4 \times 10^{-11}}$& 262 &${8\times 10^{6}}$ \\ 
& 59886.1846954935(5) & $1.0\times 10^{-3}$ & \nicer & $> 0.12$ & ${4.4 \times 10^{-11}}$& 262 &${5\times 10^{8}}$ \\ 
& 59886.1892703034(5) & $1.6\times 10^{-2}$ & \nicer & $> 0.85$ & ${4.4 \times 10^{-11}}$& 262 &${8\times 10^{7}}$ \\ 
& 59889.1707981026(5) & $3.2\times 10^{-2}$ & \nicer & $> 2.02$ & ${4.4 \times 10^{-11}}$& 262 &${3\times 10^{7}}$ \\ 
& 59889.1720946333(5) & $2.0\times 10^{-2}$ & \nicer & $> 0.66$ & ${4.4 \times 10^{-11}}$& 262 &${1\times 10^{8}}$ \\ 
& 59889.1738768160(5) & $2.6\times 10^{-2}$ & \nicer & $> 0.65$ & ${4.4 \times 10^{-11}}$& 262 &${1\times 10^{8}}$ \\ 
& 59889.1742898498(5) & $1.2\times 10^{-2}$ & \nicer & $13.6(15)$ & ${4.4 \times 10^{-11}}$& 262 &${8\times 10^{6}}$ \\ 
& 59889.1748964928(5) & $2.0\times 10^{-2}$ & \nicer & $> 0.54$ & ${4.4 \times 10^{-11}}$& 262 &${1\times 10^{8}}$ \\ 
& 59889.1749662400(5) & $2.4\times 10^{-2}$ & \nicer & $> 1.69$ & ${4.4 \times 10^{-11}}$& 262 &${4\times 10^{7}}$ \\ 
&  59280.80173402587& $1.7 \times 10^{-3}$&\xmm & $12.6(4)$& ${2.6\times 10^{-11}}$ & 180 & ${7\times 10^6}$\\ 
\enddata
\end{deluxetable*}
\tablenotetext{a}{reported $\ate$ values correspond to the 0.997 upper limit}
\end{longrotatetable}
\end{document}

%% file: authors.tex
\author[0000-0001-6422-8125]{Amanda M. Cook}
\affiliation{Dunlap Institute for Astronomy \& Astrophysics, University of Toronto, 50 St. George Street, Toronto, Ontario, Canada M5S 3H4}
\affiliation{David A. Dunlap Institute Department of Astronomy \& Astrophysics, University of Toronto, 50 St. George Street, Toronto, Ontario, Canada M5S 3H4}

\author[0000-0002-7374-7119]{Paul Scholz}
\affiliation{Department of Physics and Astronomy, York University, 4700 Keele Street, Toronto, ON MJ3 1P3, Canada}
\affiliation{Dunlap Institute for Astronomy \& Astrophysics, University of Toronto, 50 St. George Street, Toronto, Ontario, Canada M5S 3H4}

\author[0000-0002-8912-0732]{Aaron B. Pearlman}
\altaffiliation{Banting Fellow, McGill Space Institute~(MSI) Fellow, \\ and FRQNT Postdoctoral Fellow.}
\affiliation{Department of Physics, McGill University, 3600 rue University, Montr\'eal, QC H3A 2T8, Canada}
\affiliation{Trottier Space Institute, McGill University, 3550 rue University, Montr\'eal, QC H3A 2A7, Canada}

\author[0000-0001-5002-0868]{Thomas C. Abbott}
\affiliation{Department of Physics, McGill University, 3600 rue University, Montr\'eal, QC H3A 2T8, Canada}
\affiliation{Trottier Space Institute, McGill University, 3550 rue University, Montr\'eal, QC H3A 2A7, Canada}
\author[0000-0001-6804-6513]{Marilyn Cruces}
\affiliation{European Southern Observatory, Karl-Schwarzschild-Str. 2,
85748 Garching bei M\"unchen,
Germany}
\affiliation{Joint ALMA Observatory, Alonso de Córdova 3107, Vitacura, Santiago, Chile}
\affiliation{Max-Planck-Institut für Radioastronomie, Auf dem Hügel 69, 53121 Bonn, Germany}
\affiliation{Centre of Astro-Engineering, Pontificia Universidad Catolica de Chile, Av. Vicu\~{n}a Mackenna 4860, Santiago, Chile}
\affiliation{Department of Electrical Engineering, Pontificia Universidad Catolica de Chile, Av. Vicu\~{n}a Mackenna 4860, Santiago, Chile}
\author[0000-0002-3382-9558]{B. M. Gaensler}
\affiliation{Dunlap Institute for Astronomy \& Astrophysics, University of Toronto, 50 St. George Street, Toronto, Ontario, Canada M5S 3H4}
\affiliation{David A. Dunlap Institute Department of Astronomy \& Astrophysics, University of Toronto, 50 St. George Street, Toronto, Ontario, Canada M5S 3H4}
\affiliation{Department of Astronomy and Astrophysics, University of California Santa Cruz, 1156 High Street, Santa Cruz, CA 95064, USA}

\author[0000-0003-4098-5222]{Fengqiu Adam Dong}
\affiliation{Department of Physics and Astronomy, University of British Columbia, 6224 Agricultural Road, Vancouver, BC V6T 1Z1 Canada}
\author[0000-0002-2551-7554]{Daniele Michilli}
\affiliation{MIT Kavli Institute for Astrophysics and Space Research, Massachusetts Institute of Technology, 77 Massachusetts Ave, Cambridge, MA 02139, USA}
\affiliation{Department of Physics, Massachusetts Institute of Technology, 77 Massachusetts Ave, Cambridge, MA 02139, USA}
\author[0000-0003-3734-8177]{Gwendolyn Eadie}
\affiliation{David A. Dunlap Institute Department of Astronomy \& Astrophysics, University of Toronto, 50 St. George Street, Toronto, Ontario, Canada M5S 3H4}
\affiliation{Department of Statistical Science, University of Toronto, Ontario Power Building, 700 University Avenue, 9th Floor, Toronto, ON, Canada M5G 1Z5}
\affiliation{Data Sciences Institute, University of Toronto, 700 University Avenue, 10th Floor, Toronto, ON M5G 1Z5}
\author[0000-0001-9345-0307]{Victoria M. Kaspi}
\affiliation{Department of Physics, McGill University, 3600 rue University, Montr\'eal, QC H3A 2T8, Canada}
\affiliation{Trottier Space Institute, McGill University, 3550 rue University, Montr\'eal, QC H3A 2A7, Canada}
\author[0000-0001-9784-8670]{Ingrid Stairs}
\affiliation{Department of Physics and Astronomy, University of British Columbia, 6224 Agricultural Road, Vancouver, BC V6T 1Z1 Canada}
\author[0000-0001-7509-0117]{Chia Min Tan}
\affiliation{International Centre for Radio Astronomy Research, Curtin University, Bentley, WA 6102, Australia}
\author[0000-0002-3615-3514]{Mohit Bhardwaj}
\affiliation{McWilliams Center for Cosmology and Astrophysics, Department of Physics, Carnegie Mellon University, Pittsburgh, PA 15213, USA}
\author[0000-0003-2047-5276]{Tomas Cassanelli}
\affiliation{Department of Electrical Engineering, Universidad de Chile, Av. Tupper 2007, Santiago 8370451, Chile}


\author[0000-0002-8376-1563]{Alice P. Curtin}
\affiliation{Department of Physics, McGill University, 3600 rue University, Montr\'eal, QC H3A 2T8, Canada}
\affiliation{Trottier Space Institute, McGill University, 3550 rue University, Montr\'eal, QC H3A 2A7, Canada}
\author[0000-0003-2405-2967]{Adaeze L. Ibik}
\affiliation{Dunlap Institute for Astronomy \& Astrophysics, University of Toronto, 50 St. George Street, Toronto, Ontario, Canada M5S 3H4}
\affiliation{David A. Dunlap Institute Department of Astronomy \& Astrophysics, University of Toronto, 50 St. George Street, Toronto, Ontario, Canada M5S 3H4}
\author[0000-0002-5857-4264]{Mattias Lazda}
\affiliation{David A. Dunlap Institute Department of Astronomy \& Astrophysics, University of Toronto, 50 St. George Street, Toronto, Ontario, Canada M5S 3H4}
\affiliation{Dunlap Institute for Astronomy \& Astrophysics, University of Toronto, 50 St. George Street, Toronto, Ontario, Canada M5S 3H4}

\author[0000-0002-4279-6946]{Kiyoshi W. Masui}
\affiliation{MIT Kavli Institute for Astrophysics and Space Research, Massachusetts Institute of Technology, 77 Massachusetts Ave, Cambridge, MA 02139, USA}
\affiliation{Department of Physics, Massachusetts Institute of Technology, 77 Massachusetts Ave, Cambridge, MA 02139, USA}
\author[0000-0002-8897-1973]{Ayush Pandhi}
\affiliation{Dunlap Institute for Astronomy \& Astrophysics, University of Toronto, 50 St. George Street, Toronto, Ontario, Canada M5S 3H4}
\affiliation{David A. Dunlap Institute Department of Astronomy \& Astrophysics, University of Toronto, 50 St. George Street, Toronto, Ontario, Canada M5S 3H4}

\author[0000-0001-7694-6650]{Masoud Rafiei-Ravandi}
\affiliation{Department of Physics, McGill University, 3600 rue University, Montr\'eal, QC H3A 2T8, Canada}
\affiliation{Trottier Space Institute, McGill University, 3550 rue University, Montr\'eal, QC H3A 2A7, Canada}

\author[0000-0002-4623-5329]{Mawson W. Sammons}
\affiliation{Trottier Space Institute, McGill University, 3550 rue University, Montr\'eal, QC H3A 2A7, Canada}
\author[0000-0002-6823-2073]{Kaitlyn Shin}
\affiliation{MIT Kavli Institute for Astrophysics and Space Research, Massachusetts Institute of Technology, 77 Massachusetts Ave, Cambridge, MA 02139, USA}
\affiliation{Department of Physics, Massachusetts Institute of Technology, 77 Massachusetts Ave, Cambridge, MA 02139, USA}

\author[0000-0002-2088-3125]{Kendrick Smith}
\affiliation{Perimeter Institute for Theoretical Physics, 31 Caroline Street N, Waterloo, ON N25 2YL, Canada}

\author[0000-0002-9761-4353]{David C. Stenning}
\affiliation{Department of Statistics \& Actuarial Science, Simon Fraser University, Burnaby, BC, Canada}